\documentclass[sigconf,nonacm]{acmart}
\pdfoutput=1

\usepackage{amsmath}
\usepackage{algorithmic}
\usepackage{graphicx}
\usepackage{textcomp}
\usepackage{xcolor}

\usepackage{pgfplots}
\usepackage{pgfplotstable}
\pgfplotsset{compat=newest}
\usepackage{algorithm}
\usepackage{makecell}
\usepackage{enumitem}

\usepackage{tikz}
\usetikzlibrary{arrows,shapes,trees}
\usetikzlibrary{arrows.meta}

\AtBeginDocument{}

\newcommand\blfootnote[1]{%
  \begingroup
  \renewcommand\thefootnote{}\footnote{#1}%
  \addtocounter{footnote}{-1}%
  \endgroup
}

\begin{document}

\title{Efficient Multi-Processor Scheduling in Increasingly Realistic Models}

\author{P\'al Andr\'as Papp}
\orcid{0009-0005-6667-802X}
\email{pal.andras.papp@huawei.com}
\affiliation{
  \department{Computing Systems Lab}
  \institution{Huawei Zurich Research Center}
  \city{Zurich}
  \country{Switzerland}
}

\author{Georg Anegg}
\email{georg.anegg@huawei.com}
\orcid{0000-0002-5730-5812}
\affiliation{
  \department{Computing Systems Lab}
  \institution{Huawei Zurich Research Center}
  \city{Zurich}
  \country{Switzerland}
}

\author{Aikaterini Karanasiou}
\email{aikaterini.karanasiou@huawei.com}
\orcid{0000-0002-6152-863X}
\affiliation{
  \department{Computing Systems Lab}
  \institution{Huawei Zurich Research Center}
  \city{Zurich}
  \country{Switzerland}
}

\author{Albert-Jan N. Yzelman}
\email{albertjan.yzelman@huawei.com}
\orcid{0000-0001-8842-3689}
\affiliation{
 \department{Computing Systems Lab}
  \institution{Huawei Zurich Research Center}
  \city{Zurich}
  \country{Switzerland}
}

\begin{abstract}
We study the problem of efficiently scheduling a computational DAG on multiple processors. The majority of previous works have developed and compared algorithms for this problem in relatively simple models; in contrast to this, we analyze this problem in a more realistic model that captures many real-world aspects, such as communication costs, synchronization costs, and the hierarchical structure of modern processing architectures. For this we extend the well-established BSP model of parallel computing with non-uniform memory access (NUMA) effects. We then develop a range of new scheduling algorithms to minimize the scheduling cost in this more complex setting: several initialization heuristics, a hill-climbing local search method, and several approaches that formulate (and solve) the scheduling problem as an Integer Linear Program (ILP). We combine these algorithms into a single framework, and conduct experiments on a diverse set of real-world computational DAGs to show that the resulting scheduler significantly outperforms both academic and practical baselines. In particular, even without NUMA effects, our scheduler finds solutions of $24\%-44\%$ smaller cost on average than the baselines, and in case of NUMA effects, it achieves up to a factor $2.5\times$ improvement compared to the baselines. Finally, we also develop a multilevel scheduling algorithm, which provides up to almost a factor $5\times$ improvement in the special case when the problem is dominated by very high communication costs.
\end{abstract}




\maketitle

\blfootnote{\copyright P\'al Andr\'as Papp, Georg Anegg, Aikaterini Karanasiou and Albert-Jan N. Yzelman, 2024. This is the author's full version of the work, posted here for personal use. Not for redistribution. The definitive version (extended abstract) was published in the 36th ACM Symposium on Parallelism in Algorithms and Architectures (SPAA 2024), https://doi.org/10.1145/3626183.3659972.}

\section{Introduction}

The optimal scheduling of complex computational workloads on multiple processors has been extensively studied since the early days of computing. As the different subtasks in a computation usually have precedence constraints between them (one subtask can only be started after another one has been finished), the computations in these applications are often modeled as Directed Acyclic Graphs (DAGs), with the directed edges representing these dependencies.

Besides the structure of the DAG representing the computation, the other main ingredient of a scheduling problem is the machine model assumed. DAG scheduling has been studied in a very wide range of models in the past decades, differing in e.g.\ the numbers and types of processors, the modelling of communication costs, or other properties of the system such as synchronization or preemption. However, for arbitrary computational workloads (i.e.\ general DAGs), finding an optimal or close-to-optimal schedule is already very challenging in the simplest possible models, e.g.\ with uniform processors and very simple models of communication (or even no communication costs at all).

Due to this, previous research has mostly focused on variants of the scheduling problem where either the set of input DAGs or the scheduling model is heavily restricted or simplified. Theoretical work has almost exclusively studied very restricted models, where it is still manageable to analyze approximation algorithms or hardness results; however, these models are usually far from realistic. Experimental research has developed several scheduling heuristics based on natural ideas such as locality, that generally deliver good empirical results in simpler models; however, these heuristics are completely impervious to the parameters of real-world systems, and hence it is unclear how their performance carries over to more complex models. On the practical side, the parallel computing community has developed more sophisticated models, such as the Bulk Synchronous Parallel (BSP) or the LogP model, that capture many aspects of a real-world system; however, due to their complexity, these models have mostly been used for developing and analyzing parallel implementations of concrete computations (i.e.\ specific DAGs), instead of general computational tasks. As such, while each of these directions offers valuable insight, we still lack a proper understanding of how to devise efficient scheduling algorithms for arbitrary DAGs in a realistic machine model.

Our goal in this paper is to develop and analyze scheduling algorithms for this more general case: we consider the scheduling problem for general DAGs, in a model that captures many of the most important aspects of real computations. In particular, we begin from the BSP model which already accounts for communication volume and latency, and we extend this with a simple model of non-uniform memory access (NUMA) effects, which also have a defining role in modern manycore architectures. We then develop a framework of scheduling algorithms in this more realistic model, and compare this against several state-of-the-art scheduling algorithms from previous research and from applications, on a set of DAGs representing a wide range of actual computational tasks.

Our first contribution in the paper is a database of computational DAGs, which offers a diverse benchmark to evaluate different scheduling algorithms in the future. Then in terms of algorithms, we first present some heuristic methods to develop initial schedules (similar to heuristics from previous works, but tuned towards our realistic model). We then develop more advanced scheduling algorithms that take a different approach: instead of making heuristic decisions based on e.g.\ locality, they consider the actual cost of each solution in our complex model (with all its parameters), and directly optimize for minimizing this cost. In particular, we develop (i) a local search algorithm that iteratively improves a solution by making small modifications to a schedule as long as this decreases its total cost, and (ii) several different ways to model the problem as an Integer Linear Program (ILP), and use an ILP solver to find (sub)schedules of small cost. Finally, we also present a multilevel scheduling approach that is superior in the case when our scheduling problem is substantially dominated by communication costs.

Unsurprisingly, our more complex algorithms require notably more running time than lightweight heuristics, and hence their study is not viable on very large graphs. On the other hand, in the domain where they are applicable, our results show that they significantly outperform the baseline heuristics. 
In particular, even without NUMA effects, our scheduler achieves a cost reduction of $32\%$--$51\%$ compared to a simpler, and $13\%$--$40\%$ compared to a more sophisticated baseline, depending on the number of processors and the parameters of the model, on DAGs up to $10_{\,}000$ nodes. In case of NUMA effects, the improvement is even larger: $48\%$--$71\%$ to the simpler, and $27\%$--$58\%$ to the stronger baseline, or with the multilevel algorithm, even up to $79\%$ (i.e.\ almost a factor $5\times$ improvement) to the stronger baseline in some cases. This shows that in realistic scheduling models, our approach can indeed return drastically better solutions than the baseline algorithms.

\section{Related Work}

Different variants of the DAG scheduling problem have been studied thoroughly since the 1960s. In the first decades, research has mostly focused on very simple settings that either do not consider communication costs (motivated by the PRAM model), or only capture it as a fixed delay, independently of communication volume. In terms of theoretical results, scheduling is known to be already NP-hard in these simple models, and on special subclasses of DAGs \cite{PRAMcomplexity1, PRAMcomplexity2, PRAMcomplexity3, CDcomplexity1, CDcomplexity2}. On the other hand, the topic of developing approximation algorithms in these models is still extensively studied in recent years \cite{approx1, approx2, approx3, approx4}.

On the experimental side, researchers have developed a range of different scheduling heuristics, and evaluated their performance empirically. Recent surveys \cite{schedsurvey} have classified these heuristics into two groups: list-based \cite{list1, list2, list3, list4, list5} and cluster-based \cite{clus1, clus2, clus3} methods. To our knowledge, there is very little previous work on evaluating these heuristics in more realistic models (with the exception of \cite{SPD}, which is discussed separately below).

On the more practical side, BSP is one of the prominent models of parallel computing, and has been studied extensively from various perspectives \cite{BSPintro, BSPbook1, BSPbook2, BSPqa, multiBSP2}. The BSP model accounts for several important real-world aspects such as communication volume and latency, but it conveniently captures all these in a relatively simple cost function; as such, it is also used in various applications via the BSPlib library and its implementations \cite{BSPlib, BSPimpl1, BSPimpl2}. However, most of the previous works on BSP (and on similar models with more parameters, such as LogP \cite{logP}) focus on the scheduling of specific DAGs that describe a concrete computation \cite{BSPalg1, BSPalg2, BSPalg3}. One exception to this is a recent theoretical work that studies the computational complexity of BSP scheduling for given subclasses of DAGs, and also presents a naive ILP formulation to capture the BSP scheduling problem on general DAGs \cite{DAGBSP}.

The impact of NUMA in an algorithmic context has also been studied before on e.g.\ different variants of the (hyper)graph partitioning problem \cite{hier1, multi3, hyperDAG}.

Numerous other variants of the scheduling problem have also been studied extensively, e.g.\ with separate deadlines for nodes or with node duplication \cite{tardiness, duplic}.

The closest result our work is the work of Özkaya et al. \cite{SPD}, who conduct a similar experimental study with the same goal of analyzing sophisticated scheduling algorithms in a more realistic model. Our work differs from their approach in several aspects. Firstly, the work of \cite{SPD} introduces a custom model (named duplex single-port model) that only extends classical scheduling models with communication volume; in contrast to this, our work considers BSP, a well-established parallel computing model in both theory and practice which also captures further aspects such as latency, and we further extend this with NUMA effects. Moreover, while the main idea of \cite{SPD} is to extend modern list schedulers with a acyclic partitioner, we take an entirely different approach, and focus on methods that aim to directly minimize the cost function (such as local search and ILP representations).

\section{Problem definition and background}

\subsection{Preliminaries}

We assume that our computation is represented by Directed Acyclic Graph (DAG) $G(V, E)$. The nodes of the DAG correspond to subtasks or operations we need to execute, and the directed edges correspond to dependencies between these operations: an edge from node $u$ to node $v$ implies that the execution of $u$ has to be finished before the execution of $v$ begins, because the output of $u$ is required as an input for $v$. We denote the number of nodes by $n$.

Besides the structure of the DAG, our computation is described by two parameters for each node $v$: the computation weight $w(v)$ (also called \emph{work weight}) is the amount of time required to execute operation $v$ on a processor, and the \emph{communication weight} $c(v)$ is the amount of communication required to send the output of $v$ to another processor (e.g.\ its size in bytes). These weights can differ significantly between the different nodes, so they are both crucial to include in our model. Note that we consider the communication weight $c(v)$ to be a property of each node of the DAG (in contrast to e.g.\ \cite{SPD}, where it is assigned to the edges). For simplicity, we assume that node weights are integers.

\subsection{The BSP model} The BSP model assumes that the computational steps are organized into so-called \emph{supersteps}. Each superstep consists of the following two phases:
\begin{enumerate}[topsep=4pt,itemsep=0pt,partopsep=2pt,parsep=5pt]
    \item \textit{Computation phase}: each processor can execute an arbitrary amount of computation, but no communication is allowed. 
    \item \textit{Communication phase}: processors can communicate any number of values to each other, but no computation happens.
\end{enumerate}

Intuitively, supersteps correspond to larger batches of computations that are executed consecutively on a single processor, without any interruption for communication. Dividing the computations into such batches is often beneficial in practice, since the communication often comes with a significant overhead that is independent of the number of communicated values, due to e.g.\ synchronization or network initialization.

Our computing architecture in BSP is described by three parameters: the number $P$ of processors available, the time cost $g$ of sending a single unit of data between processors, and a fixed overhead cost $\ell$ (called the \emph{latency}) incurred by each superstep.

When applying the BSP model to DAG scheduling, our schedule must respect the dependencies described by the edges of the DAG. This means that a node $v$ can only be computed on processor $p$ in superstep $s$ if the output values from all its direct predecessors are already present on $p$: that is, they were either computed on $p$ in an earlier (or the same) superstep, or they were sent to $p$ by another processor before superstep $s$. In the communication phases, a processor $p$ can send the output value of any node $v$ if it is already present on $p$, to any other processor(s).

Formally, a BSP schedule of a DAG consists of (i) an assignment of nodes to processors $\pi :V_{\!} \rightarrow_{\!} \{1, \dots, P \}$ and supersteps $\tau :V _{\!}\rightarrow _{\!} \mathbb{N}$, and (ii) a communication schedule $\Gamma$, i.e.\ a set of $4$-tuples $(v, p_1, p_2, s)$, indicating that the output of node $v$ is sent from processor $p_1$ to processor $p_2$ in the communication phase of superstep $s$. A valid BSP schedule then must satisfy the conditions discussed above, i.e.\:
\begin{itemize}[topsep=4pt,itemsep=0pt,partopsep=2pt,parsep=6pt]
 \item For each edge $(u,v)$ of $G$, in case of $\pi(u)=\pi(v)$, we must have $\tau(u) \leq \tau(v)$, and in case of $\pi(u) \neq \pi(v)$, we must have an entry $(u, p_1, \pi(v), s) \in \Gamma$ for some processor $p_1$ and some superstep $s < \tau(v)$.
 \item For each $(v, p_1, p_2, s) \in \Gamma$, we must either have $\pi(v)=p_1$ and $\tau(v) \leq s$, or we must have another $(v, p', p_1, s') \in \Gamma$ with $s'<s$.
\end{itemize}

\begin{figure}
    \centering
    \resizebox{0.48\textwidth}{!}{\begin{tikzpicture}
	
    \begin{scope}[very thick, gray]

    \draw [gray, densely dotted] (0pt,55pt) -- (240pt,55pt);
    \draw [gray, densely dotted] (160pt,10pt) -- (160pt,125pt);
    \end{scope}

    \node[anchor=center, gray, rotate=90] at (-18pt,80pt) {\small \textbf{proc. 1}};

    \node[anchor=center, gray, rotate=90] at (-18pt,30pt) {\small \textbf{proc. 2}};

    \draw [gray] (0pt,108pt) -- (0pt,110pt) -- (76pt,110pt) -- (76pt,108pt);
    \draw [gray] (38pt,110pt) -- (38pt,112pt);
    \node[anchor=center, gray] at (38pt,120pt) {\small computation};

    \draw [gray] (82pt,108pt) -- (82pt,110pt) -- (156pt,110pt) -- (156pt,108pt);
    \draw [gray] (119pt,110pt) -- (119pt,112pt);
    \node[anchor=center, gray] at (119pt,120pt) {\small communication};

    \node[anchor=center, gray] at (80pt,140pt) {\small \textbf{superstep 1}};

    \draw [gray] (164pt,108pt) -- (164pt,110pt) -- (226pt,110pt) -- (226pt,108pt);
    \draw [gray] (195pt,110pt) -- (195pt,112pt);
    \node[anchor=center, gray] at (195pt,120pt) {\small computation};

    \node[anchor=center, gray] at (215pt,140pt) {\small \textbf{superstep 2}};

    \begin{scope}[thick, arrows=-stealth]
    \draw (5pt,40pt) -- (33.5pt,40pt);
    \draw (37.5pt,40pt) -- (66pt,40pt);
    \draw (5pt,20pt) -- (66pt,20pt);
    \draw (5pt,20pt) -- (35pt,37.5pt);
    \draw (37.5pt,40pt) -- (67.5pt,22.5pt);

    \draw (5pt,70pt) -- (66pt,70pt);
    \draw (5pt,70pt) -- (26.5pt,89pt);
    \draw (30pt,90pt) -- (67.5pt,72.5pt);
    \draw (30pt,90pt) -- (66pt,90pt);

    \draw (70pt,90pt) -- (166pt,90pt);
    \draw (70pt,90pt) -- (167pt,43pt);
    \draw (70pt,40pt) -- (166.5pt,87pt);
    \draw (70pt,20pt) -- (167pt,68pt);
    \draw (70pt,20pt) -- (186pt,20pt);

    \draw (170pt,40pt) -- (187.5pt,22.5pt);
    \draw (190pt,20pt) -- (216pt,20pt);
    \draw (170pt,40pt) -- (216pt,40pt);
    \draw (170pt,90pt) -- (216pt,90pt);
    \draw (170pt,70pt) -- (216.5pt,87.5pt);
    \end{scope}
    \draw[thick] (220pt,20pt) -- (230pt,20pt);
    \draw[thick] (220pt,40pt) -- (230pt,40pt);
    \draw[thick] (220pt,40pt) -- (230pt,45pt);
    \draw[thick] (220pt,90pt) -- (230pt,90pt);
    \draw[thick] (220pt,90pt) -- (230pt,85pt);

    \draw[black, fill=white] (5pt,20pt) circle (1.0ex);
    \draw[black, fill=white] (5pt,40pt) circle (1.0ex);
    \draw[black, fill=white] (37.5pt,40pt) circle (1.0ex);
    \draw[black, fill=white] (70pt,20pt) circle (1.0ex);
    \draw[black, fill=white] (70pt,40pt) circle (1.0ex);

    \draw[black, fill=white] (5pt,70pt) circle (1.0ex);
    \draw[black, fill=white] (30pt,90pt) circle (1.0ex);
    \draw[black, fill=white] (70pt,70pt) circle (1.0ex);
    \draw[black, fill=white] (70pt,90pt) circle (1.0ex);

    \draw[black, fill=white] (170pt,70pt) circle (1.0ex);
    \draw[black, fill=white] (170pt,90pt) circle (1.0ex);
    \draw[black, fill=white] (220pt,90pt) circle (1.0ex);
    \draw[black, fill=white] (170pt,40pt) circle (1.0ex);
    \draw[black, fill=white] (220pt,40pt) circle (1.0ex);
    \draw[black, fill=white] (190pt,20pt) circle (1.0ex);
    \draw[black, fill=white] (220pt,20pt) circle (1.0ex);

\end{tikzpicture}}
    \caption{Example BSP scheduling of a DAG.}
    \label{fig:BSPexample}
\end{figure}
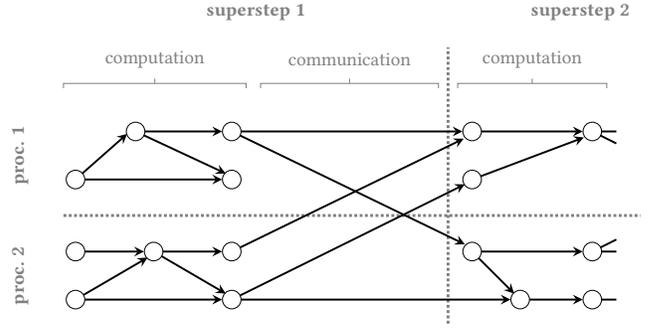

An example for a BSP scheduling of a DAG is illustrated in Figure \ref{fig:BSPexample}. In the computation phase, processors $1$ and $2$ execute $4$ and $5$ operations (nodes), respectively; then in the communication phase, processor $1$ needs to send one value to processor $2$, while processor $2$ needs to send two values to processor $1$ (to make these available on the given processors for superstep $2$). After this, the computation phase of superstep $2$ can begin.

\subsection{Cost in BSP} Another advantage of BSP is that in contrast to classical models, BSP assigns a simple cost metric to each superstep, and the total time required (i.e.\ total cost of a schedule) can be obtained by simply summing up the costs of the individual supersteps.

In the computation phase of each superstep, the subtasks assigned to each processor are executed by the processors simultaneously; as such, the cost of a computation phase is defined as the maximum work assigned to any processor in the superstep, i.e.
\[ C_{work}(s) = \max_{p \in \{1, \ldots, P\}} \, \sum_{\substack{ \pi(v)=p \\ \tau(v)=s }} \, w(v) \, . \]

On the other hand, communication costs are measured by the so-called \emph{$h$-relation} metric, where the cost of a communication phase is determined by the maximum amount of data sent or received by any processor. That is, the send cost and receive cost, respectively, of processor $p$ in superstep $s$ is
\[ C_{send}(p,s) = \sum_{\substack{(v, p_1, p_2, s_1) \in \Gamma \\ p_1=p \\ s_1=s }} \, c(v) \, , \]
\[ C_{rec}(p,s) = \sum_{\substack{(v, p_1, p_2, s_1) \in \Gamma \\ p_2=p \\ s_1=s }} \, c(v) \, , \]
and the communication cost of superstep $s$ is
\[ C_{comm}(s) = \max_{p \in \{1, \ldots, P\}} \, \max \left( C_{send}(p,s), \, C_{rec}(p,s)\right) \, . \]
The total cost of superstep $s$ is then the sum of the work, communication, and latency costs, defined as
\[ C(s) = C_{work}(s) + g \cdot C_{comm}(s) + \ell \, , \]
and the cost of the whole schedule is obtained by simply summing up $C(s)$ for all the supersteps.

For more details on the properties and the different variants of this model, we refer the reader to \cite{DAGBSP}.

\subsection{NUMA effects}

BSP already captures communications more realistically than most classical models. However, one significant drawback of BSP is that it assumes a uniform communication cost between any pair of processors. In contrast to this, today's computing architectures very often exhibit a hierarchical structure: each processor has multiple cores, each machine has multiple processors, and maybe multiple machines are connected over a network. This hierarchical structure results in a non-uniform memory access setting, where the cost of sending a piece of data heavily depends on the concrete pair of processing units that communicate: sending a value between two cores on the same processor is relatively low, whereas sending the same data over the highest level of the hierarchy (e.g.\ over the network) is drastically higher.

This asymmetry between the processors is often a defining aspect of the scheduling problem in practice. Due to this, we also extend the BSP model which such NUMA effects, and further analyze the impact of this in our experiments. The BSP definition above is straightforward to extend to this setting: if the cost $\lambda_{p_1, p_2}$ of communicating a single unit of data is known for each pair of processors $(p_1, p_2)$, then we can simply add $\lambda_{p_1, p_2}$ as a further factor in the formulas defining $C_{send}(p,s)$ and $C_{rec}(p,s)$ above. The values $\lambda_{p_1, p_2}$ then become further input parameters of the problem; we can specify them either directly for each pair of processors, or implicitly through a hierarchy. Note that the default case of uniform communication costs corresponds to a choice of $\lambda_{p_1, p_2\!}=1$ for $p_1 \neq p_2$, and $\lambda_{p_1, p_2\!}=0$ for $p_1 = p_2$.

\subsection{Problem definition}

Altogether, the input of our problem is (i) a DAG with node weights $w(v)$ and $c(v)$, and (ii) a machine description with the parameters $P$, $g$ and $\ell$, and possibly $\lambda_{p_1, p_2}$ for all pairs of processors $p_1$, $p_2$. Our goal is to find the best valid schedule in the BSP model, minimizing the total cost.

We sometimes also discuss the \emph{communication scheduling} subproblem of optimizing $\Gamma$ when $\pi$ and $\tau$ are already fixed, i.e.\ sorting the necessary communication steps into the given communication phases while minimizing the total cost of $h$-relations. With the assignment to processors and supersteps fixed, this subproblem has a much smaller degree of freedom;
its theoretical complexity was studied separately in \cite{DAGBSP}.

\section{Scheduling algorithms}

We now describe our scheduling algorithms in the BSP model. Due to space constraints, we only outline the main ideas behind each algorithm; we discuss them in more detail in Appendix \ref{app:algos}. The pipeline combining the algorithms is discussed later in Section \ref{sec:setup}.

\subsection{Baselines}

In order to evaluate our approach, we use the following baseline schedulers for comparison:

\begin{itemize}[topsep=3pt,itemsep=0pt,partopsep=2pt,parsep=6pt]
 \item \texttt{Cilk}: this is a simple and yet efficient scheduling heuristic \cite{cilk}; different variants of the same work-stealing approach are widely used in many of today's prominent parallel programming libraries and frameworks. While \texttt{Cilk} was originally not defined on DAGs, it is easy to adapt to this case. Intuitively, \texttt{Cilk}  maintains a stack of ready tasks for each processor to work on, and if a processor is idle (its stack is empty), it ``steals'' a subtask from the bottom of a (randomly chosen) other processor's stack.  We use \texttt{Cilk} to represent the baseline from the practical/application side.
 \item \texttt{BL-EST} and \texttt{ETF}: recent comparison studies have found that the best scheduling algorithms are list-based schedulers \cite{schedsurvey, SPD}. In particular, the so-called \texttt{BL-EST} and \texttt{ETF} schedulers were found to be most efficient in earlier experiments; they have also already been adapted to a setting with communication volume \cite{SPD}. When scheduling the next node, both \texttt{BL-EST} and \texttt{ETF} assign the node to the processor that offers the earliest start time, based on previous assignments and necessary communication steps. While \texttt{BL-EST} always selects the next node based on the longest outgoing path, \texttt{ETF} selects the node with the earliest starting time.
 \item \texttt{HDagg}: a more recent and more advanced state-of-the-art scheduler is the \texttt{HDagg} algorithm of Zarebavani et. al. \cite{hdagg}. \texttt{HDagg} is presented and analyzed in \cite{hdagg} for the specific purpose of speeding up SpTRSV computations; however, it is in fact a scheduling algorithm that can be applied to any computational DAG. Moreover, \texttt{HDagg} considers a very similar scheduling model to ours, sorting the nodes of the DAG into so-called wavefronts (essentially equivalent to supersteps), and minimizing the amount of communication between these wavefronts. Our experiments also show that \texttt{HDagg} consistently outperforms both \texttt{BL-EST} and \texttt{ETF}. Due to this, \texttt{HDagg} is a very fitting baseline for our work from the academic side.
\end{itemize}

Besides list-based schedulers, clustering is also a prominent method of designing state-of-the-art heuristics; however, previous work has found that this approach is consistently outperformed by \texttt{BL-EST} and \texttt{ETF} in models with communication cost \cite{SPD}.

Note that \texttt{Cilk}, \texttt{BL-EST} and \texttt{ETF} return a ``classical'' schedule where nodes are assigned to concrete time steps; such a schedule can be naturally adapted to BSP and organized into supersteps, by adding a superstep barrier (closing the current computation phase) whenever communication is required. In contrast to this, schedules returned by \texttt{HDagg} are already in the appropriate format.

\subsection{Initialization heuristics}

In order to develop an initial BSP schedule, we use the following heuristic methods:
\begin{itemize}[topsep=4pt,itemsep=0pt,partopsep=2pt,parsep=7pt]
    \item \texttt{BSPg}: A BSP-tailored greedy algorithm that consecutively assigns nodes to processors when processors become idle. Generally, we only allow assigning a node $v$ to processor $p$ if this is possible without ending the computational phase of the current superstep $s$, i.e.\ if we can ensure that all of $v$'s predecessors are already available on $p$ by superstep $s$ (that is, they were either computed on processor $p$ or in an earlier superstep). In case of multiple possible node assignments to a processor, tie-breaking is done with a heuristic that aims to minimize communication costs in the future. Once we cannot assign further nodes to at least half of the processors without a communication requirement, the computation phase of the current superstep is closed, and a next superstep is started.
    \item \texttt{Source}: A different greedy approach that in each step forms a new superstep from the next layer of source nodes in the DAG. As a preprocessing step in the beginning, the algorithm uses a simple rule to cluster the original source nodes of the DAG. Then in each superstep, it applies a round-robin-based approach to assign the current source nodes to processors, considering the nodes in a decreasing order according to $w(v)$ to ensure the load balancing of work costs between processors. Besides the current source nodes, the algorithm occasionally also adds some of their successors to the current superstep, if this requires no extra communication.
    \item \texttt{ILPinit}: We also apply an ILP-based initialization heuristic that divides the nodes into smaller batches according to a topological order, and consecutively finds a schedule for each batch separately (given the already-selected partial schedules on previous batches), using an ILP-formulation of the subproblem. This approach also provides good initial schedules, but it requires drastically more running time than the heuristics above.
\end{itemize}

Note that the above heuristics only assign the nodes to processors and supersteps, i.e.\ they only define $\pi$ and $\tau$. The required communication steps $\Gamma$ are then derived from these separately afterwards, simply following a lazy communication schedule where every required value is sent in the last possible communication phase, immediately before it is needed.

\subsection{Local search algorithm}

Another ingredient of our framework is a hill climbing local search method (denoted \texttt{HC}) that begins from an initial solution (that is, an already found, valid BSP schedule), and attempts to make small improvements to this solution as long as this is possible, i.e.\ until either a local minimum is found where none of the potential modification steps result in an improvement, or until a predefined time limit is reached.

For each local improvement step, given a node $v$ that is currently assigned to processor $p$ and superstep $s$, we consider all the alternative schedules where $v$ is assigned to any other processor $p' \neq p$ in superstep $s$, or $v$ is assigned to any processor in supersteps $(s-1)$ or $(s+1)$, with the assignments of all the other nodes unchanged. We consider all such modification steps (for every node $v$), provided that they yield a valid BSP schedule with smaller total cost.

Note this algorithm needs to be aware of the cost of the modified solution after each potential improvement step. Recalculating this entirely for each option would be very time-consuming; as such, we apply a range of sophisticated data structures to store the current schedule, and use these to efficiently query (and update) the cost change incurred by each potential improvement step, without having to consider nodes and supersteps that are unaffected by this modification.

Besides this algorithm, we also apply a separate, similar hill climbing method for the communication scheduling subproblem (denoted \texttt{HCcs}), which only tries to modify the communication schedule; that is, it checks whether any communication step $(v, p_1, p_2, s) \in \Gamma$ can be replaced by some other $(v, p_1, p_2, s')$ such that the schedule is still valid and has lower cost.

\subsection{ILP-based approach}

As our most sophisticated approach, we represent the BSP scheduling task as an ILP problem, and apply a state-of-the-art open-source ILP solver (CBC, see \cite{cbc1}) to find a low-cost schedule. We use several techniques to express (parts of) the scheduling problem as an ILP:
\begin{itemize}[topsep=2pt,itemsep=0pt,partopsep=2pt,parsep=6pt]
    \item \texttt{ILPfull}: A naive representation of BSP scheduling as an ILP problem has already been described before in the work of \cite{DAGBSP}, but only analyzed from a theoretical perspective. This formulation captures the entire problem as a single ILP, and hence it requires a very high number of variables. Due to this, even with sophisticated ILP solvers, this approach is only feasible for very small DAGs in practice.
    \item \texttt{ILPpart}: In order to handle larger DAGs, we develop a partial ILP formulation as a more advanced iterative improvement method. In particular, given a starting BSP schedule and two superstep indices $s_1 \leq s_2$, we define a partial ILP that only reorganizes the supersteps between $s_1$ and $s_2$; that is, we only consider nodes $v$ that are currently assigned to one of the supersteps in the interval $[s_1, \, s_2]$, and try to reassign these differently to any processor and any superstep in $[s_1, \, s_2]$, with the rest of the schedule unchanged. This gives a significantly smaller ILP that essentially only scales with the number of supersteps between $s_1$ and $s_2$ and the number of nodes assigned to these supersteps. Given this partial ILP formulation, we can then divide the range of supersteps into disjoint intervals, and then repeatedly use this approach to further polish each part of our schedule.
    \item \texttt{ILPcs}: We also devise and implement a separate ILP representation for the communication scheduling subproblem, i.e.\ the scheduling of communication steps $\Gamma$ when $\pi$ and $\tau$ are already fixed. This problem has a significantly smaller degree of freedom, hence the ILP solver can often return reasonably good solutions for it on the entire DAG, even for DAGs of larger size. If we already have a BSP schedule with a specific $\pi$ and $\tau$, we can use this ILP formulation to find a more optimal scheduling of the communication steps, hence resulting in a lower total cost.
\end{itemize}

When using an ILP solver in practice, we can achieve much better results by starting the solver from a good initial solution; this is provided by the initialization and local search algorithms discussed before.

\subsection{Multilevel approach}

While our algorithms above perform well in general, we have found that they are often unable to find good solutions in problems dominated by communication costs, e.g.\ when the weights $c_v$ or some of the parameters $g$, $\ell$ or $\lambda_{p_1, p_2}$ are excessively large. Intuitively speaking, in this case, a reasonable solution always needs to assign a well-connected cluster of nodes to the same processor, in order to avoid too much communication. In contrast to this, both our initialization heuristics and our local search algorithms attempt to (re)schedule single nodes separately, so they do not perform well in this case.

In order to address this problem, we also present a \emph{multilevel} scheduling algorithm, inspired by the multilevel approach that consistently provides state-of-the-art results for hypergraph partitioning and other problems \cite{multi1, multi2, multi3, DAH}. When adapted to our scheduling problem, the three main steps of the multilevel paradigm are as follows:
\begin{enumerate}[topsep=4pt,itemsep=0pt,partopsep=2pt,parsep=7pt]
    \item \textbf{Coarsening} the DAG iteratively into smaller and smaller DAGs, which (ideally) retain most of the structure of the original DAG. In our case, we do this by repeatedly contracting a selected edge $(u,v)$ of the DAG into a single node. The contracted edges are chosen in each step such that (i) the graph still remains a DAG after the contraction, and (ii) we prefer edges $(u,v)$ where the total work weight $w(u)+w(v)$ is small, but the communication weight $c(u)$ is large. 
    \item \textbf{Solving} the BSP scheduling problem in our coarsened DAG with the algorithms discussed before. The coarsened DAG is not only much smaller (and hence more viable for our algorithms), but also naturally ensures that larger clusters of nodes are assigned to the same processor and superstep.
    \item \textbf{Uncoarsening and refining} this schedule iteratively. That is, in each step, we undo the next few contraction steps in a reverse order, thus obtaining a slightly larger DAG, and we extend our current assignment $\pi$ and $\tau$ to the newly uncontracted nodes. We then execute a few iterative improvement steps (using our local search algorithm) to ensure that the schedule is further refined to fit the more delicate structure of the more uncoarsened DAG.
\end{enumerate}

Note that this approach is somewhat similar to the idea of combining an acyclic DAG partitioner with list schedulers in the work of \cite{SPD}. However, while most of the experiments in \cite{SPD} focus on creating a few (at most $4 _{\!} \cdot _{\!} P$) larger partitions, our coarsening phase sorts the DAG into a larger number of smaller clusters. Even more importantly, instead of applying a partitioning directly, our algorithm coarsens the DAG in a step-by-step fashion, since the gradual refinement steps in the uncoarsening phase are the key element of the multilevel approach.

\section{Computational DAG database}

Besides the algorithms, we also present a collection of DAGs that represent a diverse set of real-world computational tasks, and hence offer a large and realistic benchmark to evaluate our scheduling algorithms. Our computational DAG database is available at \url{https://github.com/Algebraic-Programming/HyperDAG_DB}, and its details are discussed in Appendix \ref{app:database}.

\begin{figure}
    \centering
    \begin{tikzpicture}

    \begin{scope}[thick, arrows=-stealth]
    \draw (0pt,0pt) -- (37pt,13pt);
    \draw (0pt,30pt) -- (37pt,17pt);
    \draw (120pt,0pt) -- (146pt,0pt);
    \draw (120pt,15pt) -- (146pt,15pt);
    \draw (120pt,30pt) -- (146pt,30pt);
    \draw (150pt,30pt) -- (176pt,30pt);
    \draw (150pt,0pt) -- (177pt,5.5pt);
    \draw (150pt,15pt) -- (177pt,9.5pt);
    \draw (120pt,60pt) -- (147.5pt,32.5pt);
    \draw (120pt,60pt) -- (147.5pt,17.5pt);
    \draw (120pt,45pt) -- (147.5pt,2.5pt);
    \end{scope}

    \draw[black, fill=white] (0pt,0pt) circle (1.0ex);
    \draw[black, fill=white] (0pt,30pt) circle (1.0ex);
    \draw[black, fill=white] (40pt,15pt) circle (1.0ex);
    \node[anchor=center] at (-10pt,0pt) {\small $A$};
    \node[anchor=center] at (-10pt,30pt) {\small $u$};
    \node[anchor=center] at (16pt,-20pt) {\textit{coarse-grained}};

    \draw[black, fill=white] (120pt,0pt) circle (1.0ex);
    \draw[black, fill=white] (120pt,15pt) circle (1.0ex);
    \draw[black, fill=white] (120pt,30pt) circle (1.0ex);
    \draw[black, fill=white] (120pt,45pt) circle (1.0ex);
    \draw[black, fill=white] (120pt,60pt) circle (1.0ex);
    \draw[black, fill=white] (150pt,0pt) circle (1.0ex);
    \draw[black, fill=white] (150pt,15pt) circle (1.0ex);
    \draw[black, fill=white] (150pt,30pt) circle (1.0ex);
    \draw[black, fill=white] (180pt,7.5pt) circle (1.0ex);
    \draw[black, fill=white] (180pt,30pt) circle (1.0ex);
    \node[anchor=center] at (100pt,0pt) {\small $A[2,2]$};
    \node[anchor=center] at (100pt,15pt) {\small $A[2,1]$};
    \node[anchor=center] at (100pt,30pt) {\small $A[1,1]$};
    \node[anchor=center] at (105pt,45pt) {\small $u[2]$};
    \node[anchor=center] at (105pt,60pt) {\small $u[1]$};
    \node[anchor=center] at (139pt,-20pt) {\textit{fine-grained}};

    \node[anchor=center] at (80pt,106pt) {\small example matrix-vector multiplication};
    \node[anchor=center] at (80pt,96pt) {\small (gray cells denote nonzero entries)};
    \draw[very thick, arrows=-stealth] (110pt,90pt) -- (140pt,69pt);
    \draw[very thick, arrows=-stealth] (50pt,90pt) -- (20pt,69pt);

    \draw[black, fill=gray] (45pt,120pt) rectangle (60pt,135pt);
    \draw[black, fill=gray] (75pt,120pt) rectangle (60pt,135pt);
    \draw[black, fill=gray] (45pt,150pt) rectangle (60pt,135pt);
    \draw[black, fill=white] (75pt,150pt) rectangle (60pt,135pt);
    \node[anchor=center] at (60pt,158pt) {$A$};
    
    \draw[thick, densely dashed, arrows=-stealth] (78pt,135pt) -- (95pt,135pt);
    \draw[black, fill=white] (100pt,120pt) rectangle (115pt,135pt);
    \draw[black, fill=white] (100pt,150pt) rectangle (115pt,135pt);

    \draw[thick, densely dashed, arrows=-stealth] (107.5pt,170pt) -- (107.5pt,155pt);
    \draw[black, fill=gray] (100pt,175pt) rectangle (115pt,190pt);
    \draw[black, fill=gray] (100pt,205pt) rectangle (115pt,190pt);
    \node[anchor=center] at (107.5pt,213pt) {$u$};

\end{tikzpicture}
    \caption{Coarse-grained and fine-grained DAG representation of a simple matrix-vector multiplication.}
    \label{fig:coarse_fine}
\end{figure}
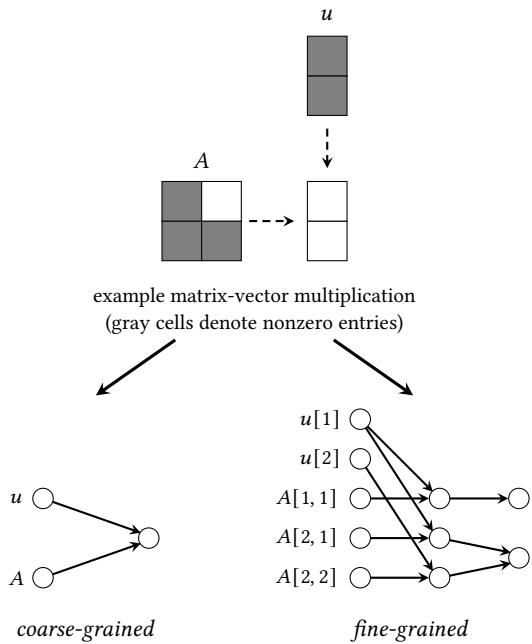

Most of the DAGs in our database correspond to computations in some algebraic form: they are derived either directly from algebraic computations, or from computations in various other areas that can naturally be expressed in algebraic form. Since these kinds of computations typically work with matrices and vectors, there are two natural ways to represent them as computational DAGs. In the \emph{coarse-grained} representation, each matrix or vector corresponds to (the output of) a single node in our DAG. In the \emph{fine-grained} representation, we also attempt to capture the internal structure of matrices/vectors: in this case, each nonzero entry in a matrix/vector is represented by (the output of) a separate node in our DAG. An example for the two representations is shown in Figure \ref{fig:coarse_fine}.

To obtain coarse-grained DAGs, we extended the C++ GraphBLAS framework from \cite{alp1} with a so-called \textit{hyperDAG backend}, which automatically extracts the structure of a given computation while the computation is running. This allows us to conveniently obtain a DAG representation for the wide variety of algorithms implemented in GraphBLAS, such as the conjugate gradient and biconjugate gradient stabilized methods, the pagerank algorithm, label propagation, $k$-nearest neighbors, and many more. Since many of these algorithms are iterative methods, we can obtain different computational DAGs with this technique if we run the given algorithms for a predefined number of iterations, or until convergence.

To obtain fine-grained DAGs, the extension of GraphBLAS would be significantly more technical, so we follow a different approach: we produce a simple tool that synthetically generates the computational DAG corresponding to four concrete algorithms (conjugate gradient, $k$-nearest neighbors, sparse matrix-vector multiplication and iterated matrix-vector multiplication) for a given pattern of nonzeros in the input matrices/vectors. While this only captures a few algorithms, it also has the advantage that we can conveniently generate a large number of computational DAGs (of any desired size) by running this generator for different matrices of a specific size and density of nonzeros.

As a technical detail, we note that in line with some recent works \cite{hyperDAG, DAH}, the DAGs in our database are represented in a hypergraph format (with hyperedges containing a node and all of its direct successors), but these are easy to convert back into a DAG format.

\section{Experimental Setup} \label{sec:setup}

For our experiments, we implemented the algorithms described above. Since the CBC ILP solver has a convenient Python interface, we implemented the ILP-based methods in Python, and the rest of the algorithms in C++. Our experiments were conducted on a workstation equipped with Intel Core i9-10900 CPU at 2.80GHz and 62.5 GiB of RAM.

The up-to-date version of our scheduling framework is available at \url{https://github.com/Algebraic-Programming/OneStopParallel}. We note that the current version of this scheduling tool has been significantly extended and upgraded since our experiments. For reproducibility, the version of the implementation used for the experiments, as well as the computational DAG database and the data from our experiments, are available in the repository at \cite{folder}.

In order to tune our algorithms, we first ran some preliminary experiments on a small training set of $10$ fine-grained computational DAGs, with their sizes ranging from $n_{\!}=_{\!}15$ to $n_{\!}=_{\!}2000$, and with BSP parameter combinations from $P _{\!} \in _{\!} \{ 4, 8, 16 \}$ and $g _{\!} \in _{\!} \{ 1, 3, 5 \}$, and a fixed $\ell=5$. These initial experiments (discussed in Appendix \ref{app:exps}) showed that \texttt{ILPinit} is only competitive when we have very few processors, while it is rather time-consuming; as such, we chose to only apply \texttt{ILPinit} for $P=4$.

We also observed that the CBC solver can usually return relatively good solutions whenever the number of variables in the ILP representation is below approximately $4_{\,}000$; we have used this as our guiding principle in \texttt{ILPpart}, choosing to extend the superstep interval $[s_1, s_2]$ until the number of variables in the resulting ILP exceeds $4_{\,}000$. Besides this, we also noted that the ILP solver rarely returns any reasonable solution above $20_{\,}000$ variables: in this case, even the preprocessing heuristics of CBC can exceed an $1$-hour time limit. Due to this, we only attempt the naive \texttt{ILPfull} approach from \cite{DAGBSP} in very small DAGs, where the estimated number of ILP variables is below $20_{\,}000$. We point out that this limitation is partially due to the fact that our work uses an open-source ILP solver; with today's more powerful commercial ILP solvers, the same approach could be applied to a significantly higher number of variables.

For our main experiments, we construct $4$ different test sets of DAGs, labeled \texttt{tiny}, \texttt{small}, \texttt{medium} and \texttt{large}, with $n$ in the ranges $[40,80]$, $[250, 500]$, $[1000, 2000]$ and $[5000, 10000]$, respectively. We generated a set of fine-grained instances with different properties (varying matrix sizes and iterations to produce some ``wider'' and some ``deeper'' DAGs), covering these intervals. This resulted in $12$ DAGs in the \texttt{tiny} dataset, and $21$ in the remaining datasets (the \texttt{tiny} set is smaller since only a few parameter options can generate so small DAGs). We further added to each of these datasets all the coarse-grained instances in the DAG database which has $n$ in the same interval.

Besides the machine parameters in the training set, we experiment with different values of $\ell$, and also with different NUMA parameters. Our NUMA settings each correspond to a binary tree hierarchy over the $P$ leaf nodes, with the communication cost increasing by a specific factor $\Delta$ over each new level, with choices of $\Delta _{\!} \in _{\!} \{2,3,4\}$. For example, in case we have $P\!=\!8$ and $\Delta\!=\!3$, then the communication costs from the first processor are $\lambda_{1, 2}\!=\!1$, then $\lambda_{1, p}\!=\!3$ for $p _{\!} \in _{\!} \{ 3, 4\}$, and $\lambda_{1, p}\!=\!9$ for $p _{\!} \in _{\!} \{ 5, 6, 7, 8\}$.

Finally, we also create a smaller dataset with DAGs of size $n \in [50000, 100000]$, called the \texttt{huge} dataset, in order to observe how our simpler algorithms scale to even larger DAGs. We do not study the ILP-based methods here, since they would be far too time-consuming in this case.

\begin{figure}
    \centering
    \begin{tikzpicture}
	
    \node[anchor=center] at (60pt,0pt) {\textit{BSP schedule}};

    \draw (40pt,35pt) rectangle (80pt,20pt);
    \node[anchor=center] at (60pt,27.5pt) {\texttt{ILPcs}};

    \draw (8pt,65pt) rectangle (112pt,50pt);
    \node[anchor=center] at (60pt,57.5pt) {\texttt{ILPfull} / \texttt{ILPpart}};

    \node[anchor=center] at (60pt,81pt) {select best};

    \draw (35pt,105pt) rectangle (85pt,120pt);
    \node[anchor=center] at (60pt,112.5pt) {\texttt{HC}$+$\texttt{HCcs}};

    \draw (-25pt,105pt) rectangle (25pt,120pt);
    \node[anchor=center] at (0pt,112.5pt) {\texttt{HC}$+$\texttt{HCcs}};

    \draw (95pt,105pt) rectangle (145pt,120pt);
    \node[anchor=center] at (120pt,112.5pt) {\texttt{HC}$+$\texttt{HCcs}};

    \draw (38pt,135pt) rectangle (82pt,150pt);
    \node[anchor=center] at (60pt,142.5pt) {\texttt{Source}};

    \draw (-18pt,135pt) rectangle (18pt,150pt);
    \node[anchor=center] at (0pt,142pt) {\texttt{BSPg}};

    \draw (96pt,135pt) rectangle (144pt,150pt);
    \node[anchor=center] at (120pt,142.5pt) {\texttt{ILPinit}};

    \node[anchor=center] at (60pt,189pt) {\textit{scheduling problem}};
    \node[anchor=center] at (60pt,178pt) {\small (DAG$_{\,}+_{\,}$parameters)};

    \begin{scope}[thick, arrows=-stealth]
    \draw (60pt,20pt) -- (60pt,5pt);
    \draw (60pt,50pt) -- (60pt,35pt);
    \draw (60pt,76pt) -- (60pt,65pt);
    \draw (60pt,105pt) -- (60pt,85pt);
    \draw (0pt,105pt) -- (0pt,95pt) -- (52pt,95pt) -- (52pt,85pt);
    \draw (120pt,105pt) -- (120pt,95pt) -- (68pt,95pt) -- (68pt,85pt);
    \draw (0pt,135pt) -- (0pt,120pt);
    \draw (60pt,135pt) -- (60pt,120pt);
    \draw (120pt,135pt) -- (120pt,120pt);
    \draw (60pt,170pt) -- (60pt,150pt);
    \draw (52pt,170pt) -- (52pt,160pt) -- (0pt,160pt) -- (0pt,150pt);
    \draw (68pt,170pt) -- (68pt,160pt) -- (120pt,160pt) -- (120pt,150pt);
    \end{scope}

\end{tikzpicture}
    \caption{Summary of our scheduling framework.}
    \label{fig:pipeline1}
\end{figure}
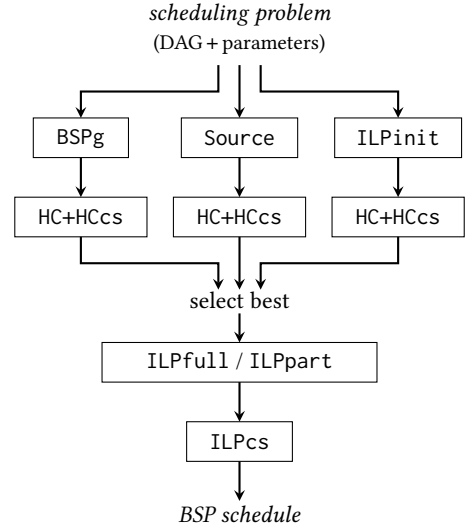

\begin{figure}
    \centering
    \hspace{32pt}
    \begin{tikzpicture}
	
    \node[anchor=center] at (60pt,0pt) {\textit{BSP schedule}};

    \draw (27pt,35pt) rectangle (93pt,20pt);
    \node[anchor=center] at (60pt,27.5pt) {\texttt{HCcs}$+$\texttt{ILPcs}};

    \draw (25pt,75pt) rectangle (95pt,50pt);
    \node[anchor=center] at (60pt,67.5pt) {uncoarsen and};
    \node[anchor=center] at (60pt,57.5pt) {refine};

    \draw (120pt,70pt) rectangle (140pt,55pt);
    \node[anchor=center] at (130pt,62.5pt) {\texttt{HC}};

    \node[anchor=center] at (60pt,100pt) {\small \textit{schedule for}};
    \node[anchor=center] at (60pt,92pt) {\small \textit{coarse DAG}};

    \draw (18pt,150pt) rectangle (102pt,120pt);
    \node[anchor=center] at (60pt,135pt) {\textbf{base scheduler}};

    \draw (28pt,180pt) rectangle (92pt,165pt);
    \node[anchor=center] at (60pt,172.5pt) {coarsen DAG};

    \node[anchor=center] at (60pt,214pt) {\textit{scheduling problem}};
    \node[anchor=center] at (60pt,203pt) {\small (DAG$_{\,}+_{\,}$parameters)};

    \begin{scope}[thick, arrows=-stealth]
    \draw (60pt,20pt) -- (60pt,5pt);
    \draw (60pt,50pt) -- (60pt,35pt);
    \draw (60pt,87pt) -- (60pt,75pt);
    \draw (60pt,120pt) -- (60pt,106pt);
    \draw (60pt,165pt) -- (60pt,150pt);
    \draw (60pt,195pt) -- (60pt,180pt);
    \draw[densely dashed] (120pt,62.5pt) -- (95pt,62.5pt);
    \end{scope}

\end{tikzpicture}
    \caption{Summary of our multilevel framework. Base scheduler refers to the pipeline in Figure \ref{fig:pipeline1} (without \texttt{ILPcs}).}
    \label{fig:pipeline2}
\end{figure}
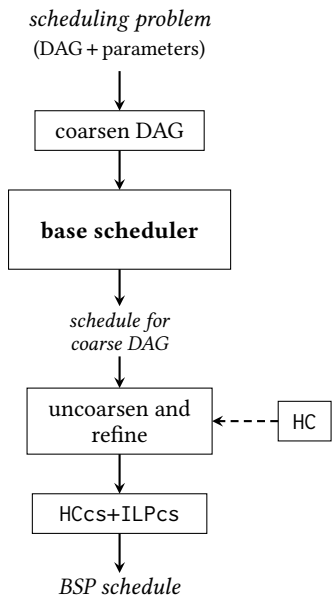

For the experiments, we combine our algorithms in the pipeline shown in Figure \ref{fig:pipeline1}. We begin by running the initialization heuristics to obtain different initial schedules. We then improve each of these separately with the local search methods (since running \texttt{HC}+\texttt{HCcs} is rather inexpensive), and then select the best schedule obtained this way. We then use the ILP-based methods: we begin with \texttt{ILPfull} in the few cases where the number of variables is below $20_{\,}000$. If \texttt{ILPfull} was not applicable, or its solution was not shown to be optimal by the ILP solver, we apply the further ILP-based methods \texttt{ILPpart} and \texttt{ILPCs}.

When \texttt{ILPfull} is applicable, we assign a time limit of $1$ hour to this. We allow $5$ minutes for \texttt{HC}+\texttt{HCcs} and $5$ minutes for \texttt{ILPcs} (although these algorithms rarely time out), and $3$ minutes for each iteration of \texttt{ILPpart}.

When using the multilevel approach, the pipeline begins with a coarsening step (see Figure \ref{fig:pipeline2}). We then apply the scheduling framework shown in Figure \ref{fig:pipeline1}, and finally refine the DAG in the end. Since the coarsened versions of the DAGs only provide an imprecise estimation of the real amount of communication required in the original DAG, we explicitly run the communication scheduling algorithms (\texttt{HCcs} and \texttt{ILPcs}) after the uncoarsening phase.

\section{Empirical Results} \label{sec:exp}

On any given problem instance, we evaluate the scheduling algorithms by the ratio between the cost returned by our algorithm and the baselines. In each dataset/experiment, we then aggregate these ratios through a geometric mean (this is more accurate for ratios than the average), to obtain a metric of the overall improvement achieved by our algorithms (with respect to the baselines) on the dataset. The results in our figures are normalized with respect to the cost of the \texttt{Cilk} baseline.

\begin{table*}[ht]
    \centering
      \caption{Results achieved by our scheduler (in a NUMA-free setting), restricted to given values of $g$, $P$ and given datasets. The two numbers in each cell show the reduction in cost compared to \texttt{Cilk} and \texttt{HDagg}, respectively.}
  \raisebox{8pt}{
  \hspace{0.05\textwidth}
  \begin{minipage}[b]{0.45\textwidth}
    \renewcommand{\arraystretch}{1.55}
    \begin{tabular}{c || c | c | c|}
      & $g=1$ & $g=3$ & $g=5$ \\ [0.5ex] 
     \hline\hline
     $P=4$ & $32\% \: / \: 20\%$  & $40\% \: / \: 24\%$ & $44\% \: / \: 26\%$ \\ 
     \hline
     $P=8$ & $40\% \: / \: 21\%$ & $45\% \: / \: 25\%$ & $45\% \: / \: 25\%$ \\
     \hline
     $P=16$ & $43\% \: / \: 18\%$ & $49\% \: / \: 26\%$ & $51\% \: / \: 30\%$ \\
     \hline
    \end{tabular}
  \end{minipage}}
  \hspace{0.01\textwidth}
    \begin{minipage}[b]{0.45\textwidth}
    \renewcommand{\arraystretch}{1.55}
    \begin{tabular}{c || c | c | c| }
     & $g=1$ & $g=3$ & $g=5$ \\ [0.5ex] 
     \hline\hline
     \texttt{tiny} & $32\% \: / \: 26\%$  & $40\% \: / \: 35\%$ & $43\% \: / \: 40\%$ \\ 
     \hline
     \texttt{small} & $38\% \: / \: 22\%$ & $44\% \: / \: 30\%$ & $46\% \: / \: 32\%$ \\
     \hline
     \texttt{medium} & $43\% \: / \: 17\%$ & $47\% \: / \: 21\%$ & $49\% \: / \: 23\%$ \\
     \hline
     \texttt{large} & $41\% \: / \: 13\%$ & $46\% \: / \: 14\%$ & $48\% \: / \: 15\%$ \\
     \hline
    \end{tabular}
  \end{minipage}
  \hfill
  \label{tab:base}
\end{table*}

As for our remaining baselines, we found that \texttt{ETF} and \texttt{BL-EST} are consistently outperformed by \texttt{HDagg}, so out of these, we only include \texttt{HDagg} in our figures and tables. The schedules achieved by \texttt{ETF} and \texttt{BL-EST} are briefly discussed in Appendix \ref{app:exps}.

For simplicity, besides the baselines, our diagrams only show the best initialization method (labelled \texttt{Init}), the cost after applying \texttt{HC}+\texttt{HCcs} (labelled \texttt{HCcs}), and the final cost after all the ILP methods (labelled \texttt{ILP}). More details on the results (e.g.\ improvements by each separate ILP method) are also provided in Appendix \ref{app:exps}.

\subsection{Without NUMA} \label{sec:nonnumaexp}

We first run our schedulers on the \texttt{tiny}, \texttt{small}, \texttt{medium} and \texttt{large} datasets, for $P _{\!} \in _{\!} \{ 4, 8, 16 \}$, $g _{\!} \in _{\!} \{1, 3, 5\}$ and $\ell\!=\!5$, without NUMA. Over the combination of all parameters and datasets, the mean cost ratio between \texttt{Cilk} and our scheduler is $0.56$, while the cost ratio between \texttt{HDagg} and our scheduler is $0.76$. This means that our approach indeed returns schedules with significantly lower cost than the baselines: it achieves a $44\%$ cost reduction compared to \texttt{Cilk}, and a $24\%$ reduction compared to \texttt{HDagg}.

A detailed analysis also shows that this improvement factor depends on the choice of parameters and dataset. In particular, Table~\ref{tab:base} shows the cost reduction with respect to the baselines, separated according to $g$ and $P$ (left), and $g$ and the dataset (right). The table shows that the improvement ranges from $32\%$ to $51\%$ compared to \texttt{Cilk}, and from $13\%$ to $40\%$ compared to \texttt{HDagg}. We can observe that the difference between our scheduler and the baselines consistently grows larger with higher $g$; this is because communication costs become more dominant, but these are not (accurately) considered by the baselines. The same holds for larger $P$ values, since this often result in more data that needs to be communicated in a schedule. As for the size of the dataset, the tables suggest that the improvement compared to \texttt{Cilk} is mostly unaffected, while the improvement compared to \texttt{HDagg} becomes smaller for large datasets. This is likely because our schedulers come with a higher time complexity, and hence they do not scale as well as \texttt{HDagg}; improving this property is a strong candidate for our future work. However, we note that even in this cases, our schedules are still consistently better than those returned by \texttt{HDagg}.

\begin{figure*}[ht]
    \centering
  \begin{minipage}[b]{0.3\textwidth}
    \resizebox{1.0\textwidth}{!}{
    \begin{tikzpicture} 
    \begin{axis}
        [ybar,
        ymin=0,ymax=1.05,
        bar width=15pt,
        height=6cm, width=7cm,
        enlarge x limits={abs=20pt},
        symbolic x coords={\texttt{Cilk}, \texttt{HDagg}, \texttt{Init}, \texttt{HCcs}, \texttt{ILP}},
        xlabel={$g=1$},
        major x tick style = transparent]
    \addplot[draw=black, fill=gray] coordinates {
        (\texttt{Cilk},1) 
        (\texttt{HDagg},0.758) 
        (\texttt{Init},0.687) 
        (\texttt{HCcs},0.635)
        (\texttt{ILP},0.61)
    };
    \end{axis}
    \end{tikzpicture}}
  \end{minipage}
  \hspace{0.02\textwidth}
    \begin{minipage}[b]{0.3\textwidth}
    \resizebox{1.0\textwidth}{!}{
    \begin{tikzpicture} 
    \begin{axis}
        [ybar,
        ymin=0,ymax=1.05,
        bar width=15pt,
        height=6cm, width=7cm,
        enlarge x limits={abs=20pt},
        symbolic x coords={\texttt{Cilk}, \texttt{HDagg}, \texttt{Init}, \texttt{HCcs}, \texttt{ILP}},
        xlabel={$g=3$},
        major x tick style = transparent]
    \addplot[draw=black, fill=gray] coordinates {
        (\texttt{Cilk},1) 
        (\texttt{HDagg},0.738) 
        (\texttt{Init},0.652) 
        (\texttt{HCcs},0.588)
        (\texttt{ILP},0.568)
    };
    \end{axis}
    \end{tikzpicture}}
  \end{minipage}
  \hspace{0.02\textwidth}
    \begin{minipage}[b]{0.3\textwidth}
    \resizebox{1.0\textwidth}{!}{
    \begin{tikzpicture} 
    \begin{axis}
        [ybar,
        ymin=0,ymax=1.05,
        bar width=15pt,
        height=6cm, width=7cm,
        enlarge x limits={abs=20pt},
        symbolic x coords={\texttt{Cilk}, \texttt{HDagg}, \texttt{Init}, \texttt{HCcs}, \texttt{ILP}},
        xlabel={$g=5$},
        major x tick style = transparent]
    \addplot[draw=black, fill=gray] coordinates {
        (\texttt{Cilk},1) 
        (\texttt{HDagg},0.731) 
        (\texttt{Init},0.636) 
        (\texttt{HCcs},0.569)
        (\texttt{ILP},0.531)
    };
    \end{axis}
    \end{tikzpicture}}
  \end{minipage}
  \hfill
  \caption{Performance comparison of \texttt{Cilk}, \texttt{HDagg} and our scheduling algorithms without NUMA effects, for values $g \in \{1,3,5\}$.}
  \label{fig:base_diag}
\end{figure*}
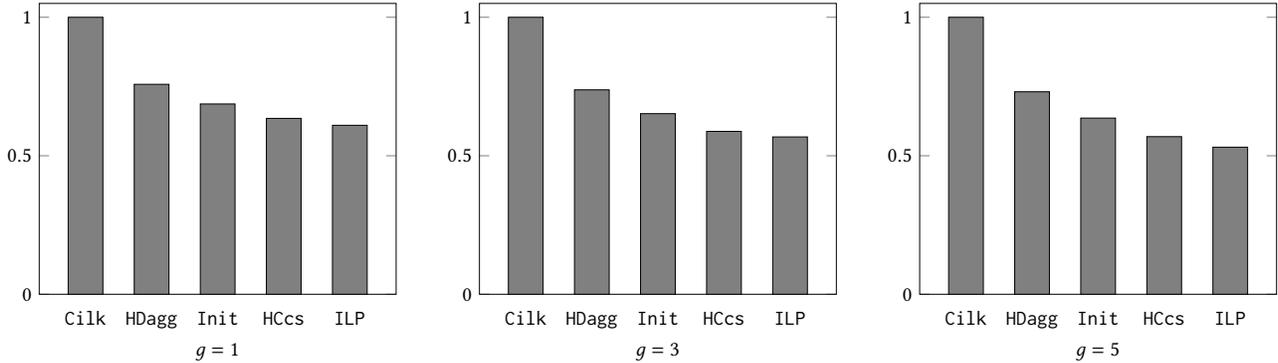

In order to understand the role of each of our different algorithmic ingredients, we also show the cost ratios achieved by the algorithms (compared to \texttt{Cilk}), separated for different values of $g$, in Figure \ref{fig:base_diag}. The figure shows that while the initialization heuristics are already much more efficient than the baselines, the local search and ILP methods both achieve some further improvement. In Appendix \ref{app:exps}, we also discuss the role of the different algorithms in more detail; for instance, the ILP-based methods tend to result in a more significant improvement on the smaller datasets, and only a minor improvement for larger DAGs.

We also considered the effect of the latency parameter $\ell$ on our schedules separately (see Appendix \ref{app:exps} for details). Our experiments show that the difference between our scheduler and the baselines also grows for larger values of $\ell$, although not as rapidly and as consistently as for the parameter $g$. This is in line with our previous observations, since the latency $\ell$ is essentially a different kind of communication cost, which is once again not considered by the baseline methods. 

For our preliminary experiments on the \texttt{huge} dataset, we only applied the non-ILP methods from our algorithmic framework: \texttt{BSPg} and \texttt{Source}, followed by \texttt{HC}+\texttt{HCcs} with a larger time limit of $30$ minutes. Even for this larger dataset, our scheduler achieves an improvement ranging from $15\%$ to $41\%$ compared to \texttt{Cilk}, and a more marginal improvement from $6\%$ to $13\%$ compared to \texttt{HDagg} (details again in Appendix \ref{app:exps}). This demonstrates that our general approach also scales reasonably well to computational DAGs with up to $100_{\,}000$ nodes.

\subsection{With NUMA effects} \label{sec:numaexp}

When we extend the BSP model with NUMA effects, the improvements achieved by our scheduler are even larger. On average over all the parameters $P _{\!} \in _{\!} \{ 8, 16 \}$ and $\Delta _{\!} \in _{\!} \{ 2,3,4 \}$, our method provides a $60\%$ improvement compared to \texttt{Cilk} and a $43\%$ improvement compared to \texttt{HDagg}.

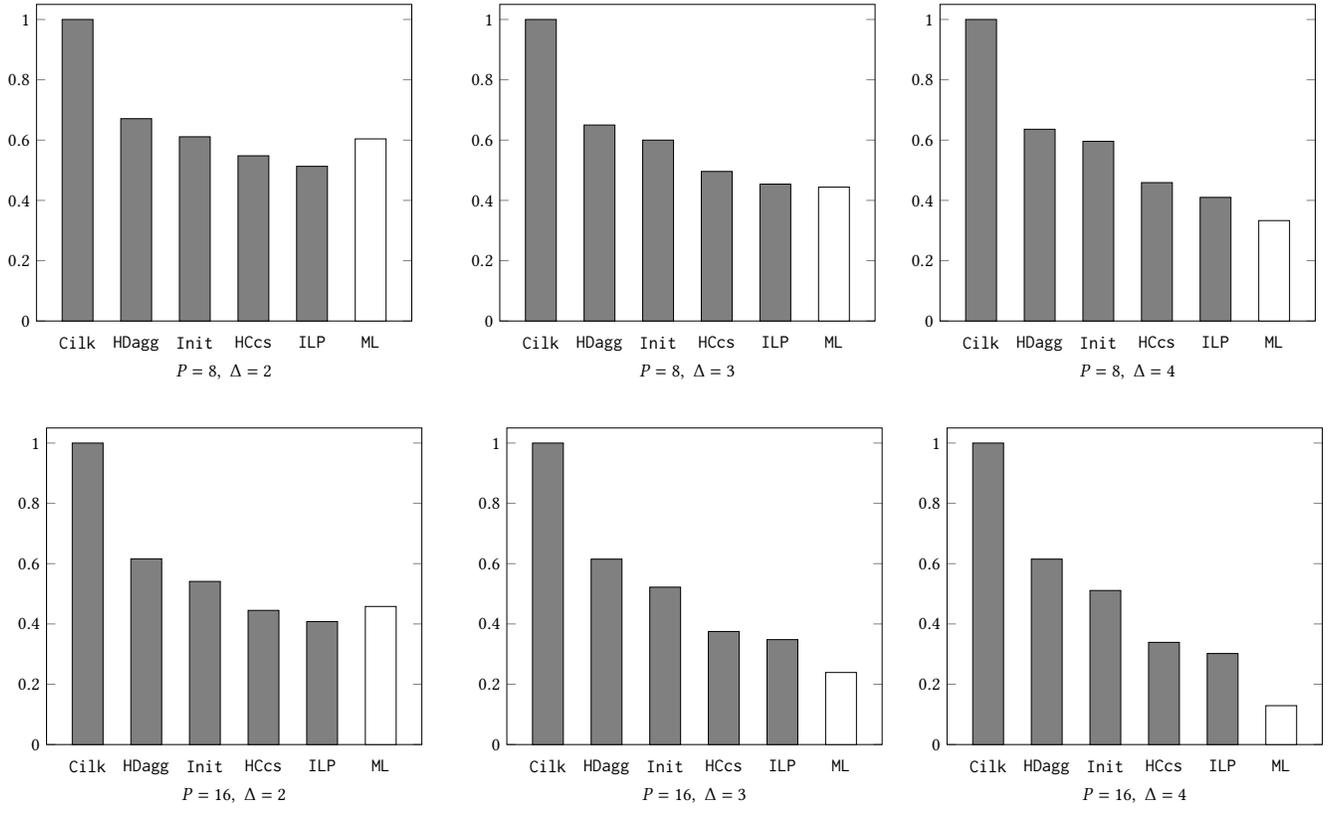
\begin{figure*}[t]
    \centering
  \begin{minipage}[b]{0.31\textwidth}
  \vspace{4pt}
    \hspace{-0.07\textwidth}
    \resizebox{1.0\textwidth}{!}{
    \begin{tikzpicture} 
    \begin{axis}
        [ybar,
        ymin=0,ymax=1.05,
        bar width=15pt,
        height=7cm, width=8cm,
        enlarge x limits={abs=20pt},
        symbolic x coords={\texttt{Cilk}, \texttt{HDagg}, \texttt{Init}, \texttt{HCcs}, \texttt{ILP}, \texttt{ML}},
        xlabel={$P=8$, $\:\Delta=2$},
        major x tick style = transparent,
        bar shift=0pt]
    \addplot[draw=black, fill=gray] coordinates {
        (\texttt{Cilk},1) 
        (\texttt{HDagg},0.671) 
        (\texttt{Init},0.611) 
        (\texttt{HCcs},0.548)
        (\texttt{ILP},0.513)
    };
    \addplot[draw=black, fill=white] coordinates {
        (\texttt{ML},0.604)
    };
    \end{axis}
    \end{tikzpicture}}
  \end{minipage}
  \hspace{0.01\textwidth}
    \begin{minipage}[b]{0.31\textwidth}
    \resizebox{1.0\textwidth}{!}{
    \begin{tikzpicture} 
    \begin{axis}
        [ybar,
        ymin=0,ymax=1.05,
        bar width=15pt,
        height=7cm, width=8cm,
        enlarge x limits={abs=20pt},
        symbolic x coords={\texttt{Cilk}, \texttt{HDagg}, \texttt{Init}, \texttt{HCcs}, \texttt{ILP}, \texttt{ML}},
        xlabel={$P=8$, $\:\Delta=3$},
        major x tick style = transparent,
        bar shift=0pt]
    \addplot[draw=black, fill=gray] coordinates {
        (\texttt{Cilk},1) 
        (\texttt{HDagg},0.65) 
        (\texttt{Init},0.6) 
        (\texttt{HCcs},0.496)
        (\texttt{ILP},0.454)
    };
    \addplot[draw=black, fill=white] coordinates {
        (\texttt{ML},0.444)
    };
    \end{axis}
    \end{tikzpicture}}
  \end{minipage}
  \hspace{0.01\textwidth}
    \begin{minipage}[b]{0.31\textwidth}
    \resizebox{1.0\textwidth}{!}{
    \begin{tikzpicture} 
    \begin{axis}
        [ybar,
        ymin=0,ymax=1.05,
        bar width=15pt,
        height=7cm, width=8cm,
        enlarge x limits={abs=20pt},
        symbolic x coords={\texttt{Cilk}, \texttt{HDagg}, \texttt{Init}, \texttt{HCcs}, \texttt{ILP}, \texttt{ML}},
        xlabel={$P=8$, $\:\Delta=4$},
        major x tick style = transparent,
        bar shift=0pt]
    \addplot[draw=black, fill=gray] coordinates {
        (\texttt{Cilk},1) 
        (\texttt{HDagg},0.636) 
        (\texttt{Init},0.596) 
        (\texttt{HCcs},0.459)
        (\texttt{ILP},0.41)
    };
    \addplot[draw=black, fill=white] coordinates {
        (\texttt{ML},0.333)
    };
    \end{axis}
    \end{tikzpicture}}
  \end{minipage}
  \hfill
  \begin{minipage}[b]{0.31\textwidth}
  \vspace{12pt}
    \resizebox{1.0\textwidth}{!}{
    \begin{tikzpicture}
    \begin{axis}
        [ybar,
        ymin=0,ymax=1.05,
        bar width=15pt,
        height=7cm, width=8cm,
        enlarge x limits={abs=20pt},
        symbolic x coords={\texttt{Cilk}, \texttt{HDagg}, \texttt{Init}, \texttt{HCcs}, \texttt{ILP}, \texttt{ML}},
        xlabel={$P=16$, $\:\Delta=2$},
        major x tick style = transparent,
        bar shift=0pt]
    \addplot[draw=black, fill=gray] coordinates {
        (\texttt{Cilk},1) 
        (\texttt{HDagg},0.616) 
        (\texttt{Init},0.541) 
        (\texttt{HCcs},0.445)
        (\texttt{ILP},0.408)
    };
    \addplot[draw=black, fill=white] coordinates {
        (\texttt{ML},0.458)
    };
    \end{axis}
    \end{tikzpicture}}
  \end{minipage}
  \hspace{0.025\textwidth}
    \begin{minipage}[b]{0.31\textwidth}
    \resizebox{1.0\textwidth}{!}{
    \begin{tikzpicture} 
    \begin{axis}
        [ybar,
        ymin=0,ymax=1.05,
        bar width=15pt,
        height=7cm, width=8cm,
        enlarge x limits={abs=20pt},
        symbolic x coords={\texttt{Cilk}, \texttt{HDagg}, \texttt{Init}, \texttt{HCcs}, \texttt{ILP}, \texttt{ML}},
        xlabel={$P=16$, $\:\Delta=3$},
        major x tick style = transparent,
        bar shift=0pt]
    \addplot[draw=black, fill=gray] coordinates {
        (\texttt{Cilk},1) 
        (\texttt{HDagg},0.615) 
        (\texttt{Init},0.522) 
        (\texttt{HCcs},0.375)
        (\texttt{ILP},0.348)
    };
    \addplot[draw=black, fill=white] coordinates {
        (\texttt{ML},0.239)
    };
    \end{axis}
    \end{tikzpicture}}
  \end{minipage}
  \hspace{0.01\textwidth}
    \begin{minipage}[b]{0.31\textwidth}
    \resizebox{1.0\textwidth}{!}{
    \begin{tikzpicture} 
    \begin{axis}
        [ybar,
        ymin=0,ymax=1.05,
        bar width=15pt,
        height=7cm, width=8cm,
        enlarge x limits={abs=20pt},
        symbolic x coords={\texttt{Cilk}, \texttt{HDagg}, \texttt{Init}, \texttt{HCcs}, \texttt{ILP}, \texttt{ML}},
        xlabel={$P=16$, $\:\Delta=4$},
        major x tick style = transparent,
        bar shift=0pt]
    \addplot[draw=black, fill=gray] coordinates {
        (\texttt{Cilk},1) 
        (\texttt{HDagg},0.615) 
        (\texttt{Init},0.511) 
        (\texttt{HCcs},0.339)
        (\texttt{ILP},0.302)
    };
    \addplot[draw=black, fill=white] coordinates {
        (\texttt{ML},0.129)
    };
    \end{axis}
    \end{tikzpicture}}
  \end{minipage}
  \hfill
  \caption{Performance comparison of \texttt{Cilk}, \texttt{HDagg} and our scheduling algorithms with NUMA effects, for different $P _{\!} \in _{\!} \{8,16\}$ and different NUMA increase factors $\Delta _{\!} \in _{\!} \{2,3,4\}$. The multilevel approach (\texttt{ML}) is shown separately at the end. This specific figure only covers the \texttt{small}, \texttt{medium} and \texttt{large} datasets, since \texttt{tiny} is too small to coarsify with \texttt{ML}; however, in general, our improvement factors in Section \ref{sec:numaexp} and Appendix \ref{app:exps} also include \texttt{tiny}.}
  \label{fig:numa_diag}
\end{figure*}

The results in this setting show an even stronger dependency on $P$ and the NUMA multiplier $\Delta$: the improvement consistently grows in both of these parameters, as also illustrated in Figure \ref{fig:numa_diag} (let us ignore the \texttt{ML} column in the figure for now). In particular, for the case of $P=8$ and $\Delta=2$, the improvement is only $48\%$ compared to \texttt{Cilk} and only $27\%$ compared to \texttt{HDagg}. On the other hand, when we have $P=16$ and $\Delta=4$, the schedule returned by our algorithm has a $71\%$ lower cost than \texttt{Cilk}, and a $58\%$ lower cost than \texttt{HDagg}. We also show the concrete improvements for each case in a numerical format in Table \ref{tab:numa2}.

Note that the case of $P=16$ and $\Delta=4$ amounts to a very significant improvement in scheduling cost: more than a $3\times$ factor for \texttt{Cilk}, and almost a $2.5\times$ factor for \texttt{HDagg}. Moreover, since $\Delta$ is often indeed large in practice, one might argue that this is the most realistic among our parameters for capturing today's computing architectures. Altogether, this suggests that our approach may indeed be able to provide significantly better schedules for many relevant applications.

We also point out that the local search and ILP methods also become much more important ingredients in this NUMA setting, achieving a notably larger further improvement than in the case without NUMA.

\subsection{Multilevel scheduling} Finally, we analyze our multilevel scheduling method, which was designed particularly for the case when communication costs are very high. For instance, in our previous setting for the highest choice of $\Delta \! = \! 4$, we have seen that our scheduler is notably better than the baselines. However, this is only their relative performance; in fact, both our scheduler and the baselines perform rather poorly in this setting, and even the solutions returned by our scheduler are often barely better (or in several cases, in fact, even worse) than the trivial solution of assigning all nodes to the same processor and superstep. Our multilevel algorithm, on the other hand, is able to find good solutions even in this rather challenging setting, consistently beating this trivial baseline.

\begin{table}[t]
\centering
\caption{Cost reduction achieved by our base scheduler in a setting with NUMA effects, for different values of $P$ and $\Delta$, compared to \texttt{Cilk} and \texttt{HDagg}, respectively.}
\renewcommand{\arraystretch}{1.65}
\begin{tabular}{c || c | c | c|}
  & $\Delta=2$ & $\Delta=3$ & $\Delta=4$ \\ [0.5ex] 
 \hline\hline
 $P=8$ & $48\% \: / \: 27\%$ & $55\% \: / \: 35\%$ & $61\% \: / \: 42\%$ \\
 \hline
 $P=16$ & $57\% \: / \: 36\%$ & $67\% \: / \: 51\%$ & $71\% \: / \: 58\%$ \\
 \hline
\end{tabular}
\label{tab:numa2}
\end{table}

The cost ratio of the multilevel method to the other schedulers is also shown in Figure \ref{fig:numa_diag} for different values of $P$ and $\Delta$. For completeness, we also show the improvement values in numerical format in Table \ref{tab:multi_small}. The figure shows that when communication costs are high (e.g.\ if we have $\Delta\!=\!4$, or if we only have $\Delta\!=\!3$ but $P\!=\!16$, and hence there are NUMA coefficients as high as $\lambda_{1, 16}\!=\!\Delta^{\log_{2\!}P-1}\!=\!27$ in our binary hierarchy), then the multilevel method is able to considerably outperform our base scheduler. For the highest choice of $P$ and $\Delta$, the multilevel algorithm finds solutions that on average provide a further factor $2\times$ improvement to our base scheduler. Altogether, this amounts to a very significant, almost a factor $5 \times$ cost reduction, even compared to \texttt{HDagg}.

\begin{table}[t]
\centering
\caption{Cost reduction achieved by the multilevel algorithm with NUMA effects, for different $P$ and $\Delta$, compared to \texttt{Cilk} and \texttt{HDagg}.}
\renewcommand{\arraystretch}{1.65}
\begin{tabular}{c || c | c | c|}
  & $\Delta=2$ & $\Delta=3$ & $\Delta=4$ \\ [0.5ex] 
 \hline\hline
 $P=8$ & $40\% \: / \: 10\%$ & $56\% \: / \: 32\%$ & $67\% \: / \: 48\%$ \\
 \hline
 $P=16$ & $54\% \: / \: 26\%$ & $76\% \: / \: 61\%$ & $87\% \: / \: 79\%$ \\
 \hline
\end{tabular}
\label{tab:multi_small}
\end{table}

However, while the multilevel algorithm achieves a large further improvement for high communication costs (large $\Delta$ and/or $P$), on the other hand, for the case of $\Delta_{\!}=_{\!}2$, the figure shows that \texttt{ML} is clearly inferior to our base scheduler. Similarly, if we revisit the setting of Section \ref{sec:nonnumaexp} without NUMA effects, the multilevel approach provides notably weaker solutions in general than our base scheduler (see Appendix \ref{app:exps} for details). This suggests that our multilevel method is indeed a specialized tool, which is useful mostly when our problem is dominated by high communication costs. However, we note that our algorithm is a relatively simple implementation of this multilevel scheduling idea; we believe that with some further work, the same approach could become an efficient tool for any choice of parameters.

\section{Conclusion}

Our results show that if we consider a more detailed and more realistic model of the scheduling problem, then applying a range of advanced algorithms can significantly outperform classical scheduling heuristics like \texttt{Cilk} or \texttt{HDagg}.

Unsurprisingly, the main drawback of these more advanced algorithms is that they come with a notably higher running time. More specifically, \texttt{BSPg} and \texttt{Source} take a similarly small amount of time as the baselines, but the remaining algorithms are more time-consuming: \texttt{HC+HCcs} and \texttt{Multi} typically take between 1-2 seconds to 1-2 minutes on our DAGs, and the ILP-based methods take even longer, hence each individual ILP is capped at a few minutes. While the running times of these initial implementations can still be reduced significantly, in general, it is inevitable that this more complex direct optimization approach requires much more time than lightweight heuristics.

However, while this running time is certainly a limitation of our work, the benefits from the higher-quality schedules can still make our approach useful in various applications, e.g.:
\begin{itemize}[topsep=3pt,itemsep=0pt,partopsep=2pt,parsep=3pt]
 \item when the same computation needs to be executed many times, possibly with different inputs,
 \item when the computation can be captured in a coarse-grained way by relatively small DAGs, but possibly still with large node weights,
 \item in areas such as machine learning, where this scheduling time may still be negligible compared to the execution of the computation itself (e.g.\ the training of a neural network).
\end{itemize}
Furthermore, even in areas where our approach is not directly applicable in practice, our algorithms still provide crucial insight on the performance of the baseline heuristics: our solutions with much smaller cost prove that the schedules returned by these baselines are often very far from the optimum. Besides this, our results also highlight the fact that many real-world aspect, such as communication volume and NUMA costs, are critical to include in our model, otherwise the algorithms may return solutions that are significantly suboptimal.

There are many natural directions for future work, i.e.\ towards finding schedules of even lower cost in BSP, or in even more detailed models. In particular, we note that most of the ingredients in our algorithmic framework are relatively simple applications of a specific technique, used as prototypes to demonstrate the feasibility of our entire approach. As such, each of these algorithmic building blocks can be further improved into, or replaced by more sophisticated methods: we could apply e.g.\ more complex local search techniques that also attempt to escape local mimima, or more efficient ILP formulations for our problems, or more advanced methods for the coarsening and refining phase of the multilevel method. We also point out that our experimental results were achieved with an open-source ILP solver; simply employing one of today's more powerful commercial ILP solvers could already significantly improve these results without any changes to the framework itself.

\begin{acks}
We would like to thank Benjamin Lozes for his valuable help with applying the \texttt{HDagg} algorithm as a baseline in our work.
\end{acks}

\bibliographystyle{ACM-Reference-Format}
\balance
\bibliography{references}


\begin{thebibliography}{46}

%%% the Web address.

\ifx \showCODEN    \undefined \def \showCODEN     #1{\unskip}     \fi
\ifx \showDOI      \undefined \def \showDOI       #1{#1}\fi
\ifx \showISBNx    \undefined \def \showISBNx     #1{\unskip}     \fi
\ifx \showISBNxiii \undefined \def \showISBNxiii  #1{\unskip}     \fi
\ifx \showISSN     \undefined \def \showISSN      #1{\unskip}     \fi
\ifx \showLCCN     \undefined \def \showLCCN      #1{\unskip}     \fi
\ifx \shownote     \undefined \def \shownote      #1{#1}          \fi
\ifx \showarticletitle \undefined \def \showarticletitle #1{#1}   \fi
\ifx \showURL      \undefined \def \showURL       {\relax}        \fi
% invisible to TeX
\providecommand\bibfield[2]{#2}
\providecommand\bibinfo[2]{#2}
\providecommand\natexlab[1]{#1}
\providecommand\showeprint[2][]{arXiv:#2}

\bibitem[fol(2024)]%
        {folder}
 \bibinfo{year}{2024}\natexlab{}.
\newblock \bibinfo{title}{Supplementary material}.
\newblock \bibinfo{howpublished}{The scheduling algorithm implementations used in our experiments, our computational DAG database, and the data from our experiments are available at: \url{https://github.com/Algebraic-Programming/Artifacts/tree/master/SPAA_2024_Efficient_Multi-Processor_Scheduling}}.
\newblock


\bibitem[Adam et~al\mbox{.}(1974)]%
        {list1}
\bibfield{author}{\bibinfo{person}{Thomas~L Adam}, \bibinfo{person}{K.~Mani Chandy}, {and} \bibinfo{person}{JR Dickson}.} \bibinfo{year}{1974}\natexlab{}.
\newblock \showarticletitle{A comparison of list schedules for parallel processing systems}.
\newblock \bibinfo{journal}{\emph{Commun. ACM}} \bibinfo{volume}{17}, \bibinfo{number}{12} (\bibinfo{year}{1974}), \bibinfo{pages}{685--690}.
\newblock


\bibitem[Ahmad and Kwok(1998)]%
        {duplic}
\bibfield{author}{\bibinfo{person}{Ishfaq Ahmad} {and} \bibinfo{person}{Yu-Kwong Kwok}.} \bibinfo{year}{1998}\natexlab{}.
\newblock \showarticletitle{On exploiting task duplication in parallel program scheduling}.
\newblock \bibinfo{journal}{\emph{IEEE Transactions on parallel and distributed systems}} \bibinfo{volume}{9}, \bibinfo{number}{9} (\bibinfo{year}{1998}), \bibinfo{pages}{872--892}.
\newblock


\bibitem[Bisseling(2020)]%
        {BSPbook2}
\bibfield{author}{\bibinfo{person}{Rob~H Bisseling}.} \bibinfo{year}{2020}\natexlab{}.
\newblock \bibinfo{booktitle}{\emph{Parallel Scientific Computation: A Structured Approach Using BSP}}.
\newblock \bibinfo{publisher}{Oxford University Press, USA}.
\newblock


\bibitem[Blumofe and Leiserson(1999)]%
        {cilk}
\bibfield{author}{\bibinfo{person}{Robert~D Blumofe} {and} \bibinfo{person}{Charles~E Leiserson}.} \bibinfo{year}{1999}\natexlab{}.
\newblock \showarticletitle{Scheduling multithreaded computations by work stealing}.
\newblock \bibinfo{journal}{\emph{Journal of the ACM (JACM)}} \bibinfo{volume}{46}, \bibinfo{number}{5} (\bibinfo{year}{1999}), \bibinfo{pages}{720--748}.
\newblock


\bibitem[Culler et~al\mbox{.}(1993)]%
        {logP}
\bibfield{author}{\bibinfo{person}{David Culler}, \bibinfo{person}{Richard Karp}, \bibinfo{person}{David Patterson}, \bibinfo{person}{Abhijit Sahay}, \bibinfo{person}{Klaus~Erik Schauser}, \bibinfo{person}{Eunice Santos}, \bibinfo{person}{Ramesh Subramonian}, {and} \bibinfo{person}{Thorsten Von~Eicken}.} \bibinfo{year}{1993}\natexlab{}.
\newblock \showarticletitle{LogP: Towards a realistic model of parallel computation}. In \bibinfo{booktitle}{\emph{Proceedings of the fourth ACM SIGPLAN symposium on Principles and practice of parallel programming}}. \bibinfo{pages}{1--12}.
\newblock


\bibitem[Forrest and Lougee-Heimer(2005)]%
        {cbc1}
\bibfield{author}{\bibinfo{person}{John Forrest} {and} \bibinfo{person}{Robin Lougee-Heimer}.} \bibinfo{year}{2005}\natexlab{}.
\newblock \showarticletitle{CBC user guide}.
\newblock In \bibinfo{booktitle}{\emph{Emerging theory, methods, and applications}}. \bibinfo{publisher}{INFORMS}, \bibinfo{pages}{257--277}.
\newblock


\bibitem[Goldman et~al\mbox{.}(1998)]%
        {BSPalg3}
\bibfield{author}{\bibinfo{person}{Alfredo Goldman}, \bibinfo{person}{Gregory Mouni{\'e}}, {and} \bibinfo{person}{Denis Trystram}.} \bibinfo{year}{1998}\natexlab{}.
\newblock \showarticletitle{Near optimal algorithms for scheduling independent chains in BSP}. In \bibinfo{booktitle}{\emph{Proceedings. Fifth International Conference on High Performance Computing}}. IEEE, \bibinfo{pages}{310--317}.
\newblock


\bibitem[Hajiaghayi et~al\mbox{.}(2014)]%
        {hier1}
\bibfield{author}{\bibinfo{person}{Mohammadtaghi Hajiaghayi}, \bibinfo{person}{Theodore Johnson}, \bibinfo{person}{Mohammad~Reza Khani}, {and} \bibinfo{person}{Barna Saha}.} \bibinfo{year}{2014}\natexlab{}.
\newblock \showarticletitle{Hierarchical graph partitioning}. In \bibinfo{booktitle}{\emph{Proceedings of the 26th ACM symposium on Parallelism in algorithms and architectures ({SPAA})}}. \bibinfo{pages}{51--60}.
\newblock


\bibitem[Herrmann et~al\mbox{.}(2017)]%
        {DAGpart}
\bibfield{author}{\bibinfo{person}{Julien Herrmann}, \bibinfo{person}{Jonathan Kho}, \bibinfo{person}{Bora U{\c{c}}ar}, \bibinfo{person}{Kamer Kaya}, {and} \bibinfo{person}{{\"U}mit~V {\c{C}}ataly{\"u}rek}.} \bibinfo{year}{2017}\natexlab{}.
\newblock \showarticletitle{Acyclic partitioning of large directed acyclic graphs}. In \bibinfo{booktitle}{\emph{2017 17th IEEE/ACM international symposium on cluster, cloud and grid computing (CCGRID)}}. IEEE, \bibinfo{pages}{371--380}.
\newblock


\bibitem[Hill et~al\mbox{.}(1998)]%
        {BSPlib}
\bibfield{author}{\bibinfo{person}{Jonathan~MD Hill}, \bibinfo{person}{Bill McColl}, \bibinfo{person}{Dan~C Stefanescu}, \bibinfo{person}{Mark~W Goudreau}, \bibinfo{person}{Kevin Lang}, \bibinfo{person}{Satish~B Rao}, \bibinfo{person}{Torsten Suel}, \bibinfo{person}{Thanasis Tsantilas}, {and} \bibinfo{person}{Rob~H Bisseling}.} \bibinfo{year}{1998}\natexlab{}.
\newblock \showarticletitle{BSPlib: The BSP programming library}.
\newblock \bibinfo{journal}{\emph{Parallel Comput.}} \bibinfo{volume}{24}, \bibinfo{number}{14} (\bibinfo{year}{1998}), \bibinfo{pages}{1947--1980}.
\newblock


\bibitem[Hoogeveen et~al\mbox{.}(1994)]%
        {CDcomplexity2}
\bibfield{author}{\bibinfo{person}{JA Hoogeveen}, \bibinfo{person}{Jan~Karel Lenstra}, {and} \bibinfo{person}{Bart Veltman}.} \bibinfo{year}{1994}\natexlab{}.
\newblock \showarticletitle{Three, four, five, six, or the complexity of scheduling with communication delays}.
\newblock \bibinfo{journal}{\emph{Operations Research Letters}} \bibinfo{volume}{16}, \bibinfo{number}{3} (\bibinfo{year}{1994}), \bibinfo{pages}{129--137}.
\newblock


\bibitem[Hwang et~al\mbox{.}(1989)]%
        {list2}
\bibfield{author}{\bibinfo{person}{Jing-Jang Hwang}, \bibinfo{person}{Yuan-Chieh Chow}, \bibinfo{person}{Frank~D Anger}, {and} \bibinfo{person}{Chung-Yee Lee}.} \bibinfo{year}{1989}\natexlab{}.
\newblock \showarticletitle{Scheduling precedence graphs in systems with interprocessor communication times}.
\newblock \bibinfo{journal}{\emph{siam journal on computing}} \bibinfo{volume}{18}, \bibinfo{number}{2} (\bibinfo{year}{1989}), \bibinfo{pages}{244--257}.
\newblock


\bibitem[Kanemitsu et~al\mbox{.}(2016)]%
        {clus3}
\bibfield{author}{\bibinfo{person}{Hidehiro Kanemitsu}, \bibinfo{person}{Masaki Hanada}, {and} \bibinfo{person}{Hidenori Nakazato}.} \bibinfo{year}{2016}\natexlab{}.
\newblock \showarticletitle{Clustering-based task scheduling in a large number of heterogeneous processors}.
\newblock \bibinfo{journal}{\emph{IEEE Transactions on Parallel and Distributed Systems}} \bibinfo{volume}{27}, \bibinfo{number}{11} (\bibinfo{year}{2016}), \bibinfo{pages}{3144--3157}.
\newblock


\bibitem[Karypis et~al\mbox{.}(1997)]%
        {multi1}
\bibfield{author}{\bibinfo{person}{George Karypis}, \bibinfo{person}{Rajat Aggarwal}, \bibinfo{person}{Vipin Kumar}, {and} \bibinfo{person}{Shashi Shekhar}.} \bibinfo{year}{1997}\natexlab{}.
\newblock \showarticletitle{Multilevel hypergraph partitioning: Application in VLSI domain}. In \bibinfo{booktitle}{\emph{Proceedings of the 34th annual Design Automation Conference}}. \bibinfo{pages}{526--529}.
\newblock


\bibitem[Kulkarni et~al\mbox{.}(2020)]%
        {approx3}
\bibfield{author}{\bibinfo{person}{Janardhan Kulkarni}, \bibinfo{person}{Shi Li}, \bibinfo{person}{Jakub Tarnawski}, {and} \bibinfo{person}{Minwei Ye}.} \bibinfo{year}{2020}\natexlab{}.
\newblock \showarticletitle{Hierarchy-based algorithms for minimizing makespan under precedence and communication constraints}. In \bibinfo{booktitle}{\emph{Proceedings of the Fourteenth Annual ACM-SIAM Symposium on Discrete Algorithms ({SODA})}}. SIAM, \bibinfo{pages}{2770--2789}.
\newblock


\bibitem[Kwok and Ahmad(1996)]%
        {clus2}
\bibfield{author}{\bibinfo{person}{Yu-Kwong Kwok} {and} \bibinfo{person}{Ishfaq Ahmad}.} \bibinfo{year}{1996}\natexlab{}.
\newblock \showarticletitle{Dynamic critical-path scheduling: An effective technique for allocating task graphs to multiprocessors}.
\newblock \bibinfo{journal}{\emph{IEEE transactions on parallel and distributed systems}} \bibinfo{volume}{7}, \bibinfo{number}{5} (\bibinfo{year}{1996}), \bibinfo{pages}{506--521}.
\newblock


\bibitem[Lenstra and Rinnooy~Kan(1978)]%
        {PRAMcomplexity1}
\bibfield{author}{\bibinfo{person}{Jan~Karel Lenstra} {and} \bibinfo{person}{AHG Rinnooy~Kan}.} \bibinfo{year}{1978}\natexlab{}.
\newblock \showarticletitle{Complexity of scheduling under precedence constraints}.
\newblock \bibinfo{journal}{\emph{Operations Research}} \bibinfo{volume}{26}, \bibinfo{number}{1} (\bibinfo{year}{1978}), \bibinfo{pages}{22--35}.
\newblock


\bibitem[Leung and Young(1990)]%
        {tardiness}
\bibfield{author}{\bibinfo{person}{Joseph Y-T Leung} {and} \bibinfo{person}{Gilbert~H Young}.} \bibinfo{year}{1990}\natexlab{}.
\newblock \showarticletitle{Minimizing total tardiness on a single machine with precedence constraints}.
\newblock \bibinfo{journal}{\emph{ORSA Journal on Computing}} \bibinfo{volume}{2}, \bibinfo{number}{4} (\bibinfo{year}{1990}), \bibinfo{pages}{346--352}.
\newblock


\bibitem[Levey and Rothvoss(2016)]%
        {approx1}
\bibfield{author}{\bibinfo{person}{Elaine Levey} {and} \bibinfo{person}{Thomas Rothvoss}.} \bibinfo{year}{2016}\natexlab{}.
\newblock \showarticletitle{A (1+ epsilon)-approximation for makespan scheduling with precedence constraints using LP hierarchies}. In \bibinfo{booktitle}{\emph{Proceedings of the forty-eighth annual ACM symposium on Theory of Computing}}. \bibinfo{pages}{168--177}.
\newblock


\bibitem[Li(2021)]%
        {approx2}
\bibfield{author}{\bibinfo{person}{Shi Li}.} \bibinfo{year}{2021}\natexlab{}.
\newblock \showarticletitle{Towards PTAS for precedence constrained scheduling via combinatorial algorithms}. In \bibinfo{booktitle}{\emph{Proceedings of the 2021 ACM-SIAM Symposium on Discrete Algorithms (SODA)}}. SIAM, \bibinfo{pages}{2991--3010}.
\newblock


\bibitem[Liu et~al\mbox{.}(2022)]%
        {approx4}
\bibfield{author}{\bibinfo{person}{Quanquan~C Liu}, \bibinfo{person}{Manish Purohit}, \bibinfo{person}{Zoya Svitkina}, \bibinfo{person}{Erik Vee}, {and} \bibinfo{person}{Joshua~R Wang}.} \bibinfo{year}{2022}\natexlab{}.
\newblock \showarticletitle{Scheduling with Communication Delay in Near-Linear Time}. In \bibinfo{booktitle}{\emph{39th International Symposium on Theoretical Aspects of Computer Science (STACS)}}.
\newblock


\bibitem[McColl(2021)]%
        {BSPbook1}
\bibfield{author}{\bibinfo{person}{Bill McColl}.} \bibinfo{year}{2021}\natexlab{}.
\newblock \showarticletitle{Mathematics, Models and Architectures}.
\newblock \bibinfo{journal}{\emph{Mathematics for Future Computing and Communications}} (\bibinfo{year}{2021}), \bibinfo{pages}{6}.
\newblock


\bibitem[McColl(1995)]%
        {BSPalg1}
\bibfield{author}{\bibinfo{person}{William~F McColl}.} \bibinfo{year}{1995}\natexlab{}.
\newblock \showarticletitle{Scalable computing}.
\newblock \bibinfo{journal}{\emph{Computer Science Today}} (\bibinfo{year}{1995}), \bibinfo{pages}{46--61}.
\newblock


\bibitem[McColl and Tiskin(1999)]%
        {BSPalg2}
\bibfield{author}{\bibinfo{person}{William~F McColl} {and} \bibinfo{person}{Alexandre Tiskin}.} \bibinfo{year}{1999}\natexlab{}.
\newblock \showarticletitle{Memory-efficient matrix multiplication in the BSP model}.
\newblock \bibinfo{journal}{\emph{Algorithmica}}  \bibinfo{volume}{24} (\bibinfo{year}{1999}), \bibinfo{pages}{287--297}.
\newblock


\bibitem[Mingsheng et~al\mbox{.}(2003)]%
        {list4}
\bibfield{author}{\bibinfo{person}{Shang Mingsheng}, \bibinfo{person}{Sun Shixin}, {and} \bibinfo{person}{Wang Qingxian}.} \bibinfo{year}{2003}\natexlab{}.
\newblock \showarticletitle{An efficient parallel scheduling algorithm of dependent task graphs}. In \bibinfo{booktitle}{\emph{4th International Conference on Parallel and Distributed Computing, Applications and Technologies}}. IEEE, \bibinfo{pages}{595--598}.
\newblock


\bibitem[{\"O}zkaya et~al\mbox{.}(2019)]%
        {SPD}
\bibfield{author}{\bibinfo{person}{M~Yusuf {\"O}zkaya}, \bibinfo{person}{Anne Benoit}, \bibinfo{person}{Bora U{\c{c}}ar}, \bibinfo{person}{Julien Herrmann}, {and} \bibinfo{person}{{\"U}mit~V {\c{C}}ataly{\"u}rek}.} \bibinfo{year}{2019}\natexlab{}.
\newblock \showarticletitle{A scalable clustering-based task scheduler for homogeneous processors using DAG partitioning}. In \bibinfo{booktitle}{\emph{IEEE International Parallel and Distributed Processing Symposium (IPDPS)}}. IEEE, \bibinfo{pages}{155--165}.
\newblock


\bibitem[Papp et~al\mbox{.}(2023a)]%
        {DAGBSP}
\bibfield{author}{\bibinfo{person}{P{\'a}l~Andr{\'a}s Papp}, \bibinfo{person}{Georg Anegg}, {and} \bibinfo{person}{AN Yzelman}.} \bibinfo{year}{2023}\natexlab{a}.
\newblock \showarticletitle{DAG Scheduling in the BSP Model}.
\newblock \bibinfo{journal}{\emph{arXiv preprint arXiv:2303.05989}} (\bibinfo{year}{2023}).
\newblock


\bibitem[Papp et~al\mbox{.}(2023b)]%
        {hyperDAG}
\bibfield{author}{\bibinfo{person}{P{\'a}l~Andr{\'a}s Papp}, \bibinfo{person}{Georg Anegg}, {and} \bibinfo{person}{Albert-Jan~N Yzelman}.} \bibinfo{year}{2023}\natexlab{b}.
\newblock \showarticletitle{Partitioning hypergraphs is hard: Models, inapproximability, and applications}. In \bibinfo{booktitle}{\emph{35th ACM Symposium on Parallelism in Algorithms and Architectures (SPAA)}}. \bibinfo{pages}{415--425}.
\newblock


\bibitem[Picouleau(1995)]%
        {CDcomplexity1}
\bibfield{author}{\bibinfo{person}{Christophe Picouleau}.} \bibinfo{year}{1995}\natexlab{}.
\newblock \showarticletitle{New complexity results on scheduling with small communication delays}.
\newblock \bibinfo{journal}{\emph{Discrete Applied Mathematics}} \bibinfo{volume}{60}, \bibinfo{number}{1-3} (\bibinfo{year}{1995}), \bibinfo{pages}{331--342}.
\newblock


\bibitem[Popp et~al\mbox{.}(2021)]%
        {DAH}
\bibfield{author}{\bibinfo{person}{Merten Popp}, \bibinfo{person}{Sebastian Schlag}, \bibinfo{person}{Christian Schulz}, {and} \bibinfo{person}{Daniel Seemaier}.} \bibinfo{year}{2021}\natexlab{}.
\newblock \showarticletitle{Multilevel Acyclic Hypergraph Partitioning}. In \bibinfo{booktitle}{\emph{2021 Proceedings of the Workshop on Algorithm Engineering and Experiments (ALENEX)}}. SIAM, \bibinfo{pages}{1--15}.
\newblock


\bibitem[Radulescu and Van~Gemund(2002)]%
        {list5}
\bibfield{author}{\bibinfo{person}{Andrei Radulescu} {and} \bibinfo{person}{Arjan~JC Van~Gemund}.} \bibinfo{year}{2002}\natexlab{}.
\newblock \showarticletitle{Low-cost task scheduling for distributed-memory machines}.
\newblock \bibinfo{journal}{\emph{IEEE transactions on parallel and distributed systems}} \bibinfo{volume}{13}, \bibinfo{number}{6} (\bibinfo{year}{2002}), \bibinfo{pages}{648--658}.
\newblock


\bibitem[Schlag et~al\mbox{.}(2016)]%
        {multi3}
\bibfield{author}{\bibinfo{person}{Sebastian Schlag}, \bibinfo{person}{Vitali Henne}, \bibinfo{person}{Tobias Heuer}, \bibinfo{person}{Henning Meyerhenke}, \bibinfo{person}{Peter Sanders}, {and} \bibinfo{person}{Christian Schulz}.} \bibinfo{year}{2016}\natexlab{}.
\newblock \showarticletitle{K-way hypergraph partitioning via n-level recursive bisection}. In \bibinfo{booktitle}{\emph{2016 Proceedings of the Eighteenth Workshop on Algorithm Engineering and Experiments (ALENEX)}}. SIAM, \bibinfo{pages}{53--67}.
\newblock


\bibitem[Skillicorn et~al\mbox{.}(1997)]%
        {BSPqa}
\bibfield{author}{\bibinfo{person}{David~B Skillicorn}, \bibinfo{person}{Jonathan Hill}, {and} \bibinfo{person}{William~F McColl}.} \bibinfo{year}{1997}\natexlab{}.
\newblock \showarticletitle{Questions and answers about BSP}.
\newblock \bibinfo{journal}{\emph{Scientific Programming}} \bibinfo{volume}{6}, \bibinfo{number}{3} (\bibinfo{year}{1997}), \bibinfo{pages}{249--274}.
\newblock


\bibitem[Svensson(2010)]%
        {PRAMcomplexity3}
\bibfield{author}{\bibinfo{person}{Ola Svensson}.} \bibinfo{year}{2010}\natexlab{}.
\newblock \showarticletitle{Conditional hardness of precedence constrained scheduling on identical machines}. In \bibinfo{booktitle}{\emph{Proceedings of the forty-second ACM symposium on Theory of computing (STOC)}}. \bibinfo{pages}{745--754}.
\newblock


\bibitem[Topcuoglu et~al\mbox{.}(2002)]%
        {list3}
\bibfield{author}{\bibinfo{person}{Haluk Topcuoglu}, \bibinfo{person}{Salim Hariri}, {and} \bibinfo{person}{Min-You Wu}.} \bibinfo{year}{2002}\natexlab{}.
\newblock \showarticletitle{Performance-effective and low-complexity task scheduling for heterogeneous computing}.
\newblock \bibinfo{journal}{\emph{IEEE transactions on parallel and distributed systems}} \bibinfo{volume}{13}, \bibinfo{number}{3} (\bibinfo{year}{2002}), \bibinfo{pages}{260--274}.
\newblock


\bibitem[Trifunovi{\'c} and Knottenbelt(2008)]%
        {multi2}
\bibfield{author}{\bibinfo{person}{Aleksandar Trifunovi{\'c}} {and} \bibinfo{person}{William~J Knottenbelt}.} \bibinfo{year}{2008}\natexlab{}.
\newblock \showarticletitle{Parallel multilevel algorithms for hypergraph partitioning}.
\newblock \bibinfo{journal}{\emph{J. Parallel and Distrib. Comput.}} \bibinfo{volume}{68}, \bibinfo{number}{5} (\bibinfo{year}{2008}), \bibinfo{pages}{563--581}.
\newblock


\bibitem[Ullman(1975)]%
        {PRAMcomplexity2}
\bibfield{author}{\bibinfo{person}{Jeffrey~D. Ullman}.} \bibinfo{year}{1975}\natexlab{}.
\newblock \showarticletitle{NP-complete scheduling problems}.
\newblock \bibinfo{journal}{\emph{Journal of Computer and System sciences}} \bibinfo{volume}{10}, \bibinfo{number}{3} (\bibinfo{year}{1975}), \bibinfo{pages}{384--393}.
\newblock


\bibitem[Valiant(1990)]%
        {BSPintro}
\bibfield{author}{\bibinfo{person}{Leslie~G Valiant}.} \bibinfo{year}{1990}\natexlab{}.
\newblock \showarticletitle{A bridging model for parallel computation}.
\newblock \bibinfo{journal}{\emph{Commun. ACM}} \bibinfo{volume}{33}, \bibinfo{number}{8} (\bibinfo{year}{1990}), \bibinfo{pages}{103--111}.
\newblock


\bibitem[Valiant(2011)]%
        {multiBSP2}
\bibfield{author}{\bibinfo{person}{Leslie~G Valiant}.} \bibinfo{year}{2011}\natexlab{}.
\newblock \showarticletitle{A bridging model for multi-core computing}.
\newblock \bibinfo{journal}{\emph{J. Comput. System Sci.}} \bibinfo{volume}{77}, \bibinfo{number}{1} (\bibinfo{year}{2011}), \bibinfo{pages}{154--166}.
\newblock


\bibitem[Wang and Sinnen(2018)]%
        {schedsurvey}
\bibfield{author}{\bibinfo{person}{Huijun Wang} {and} \bibinfo{person}{Oliver Sinnen}.} \bibinfo{year}{2018}\natexlab{}.
\newblock \showarticletitle{List-scheduling versus cluster-scheduling}.
\newblock \bibinfo{journal}{\emph{IEEE Transactions on Parallel and Distributed Systems}} \bibinfo{volume}{29}, \bibinfo{number}{8} (\bibinfo{year}{2018}), \bibinfo{pages}{1736--1749}.
\newblock


\bibitem[Yang and Gerasoulis(1994)]%
        {clus1}
\bibfield{author}{\bibinfo{person}{Tao Yang} {and} \bibinfo{person}{Apostolos Gerasoulis}.} \bibinfo{year}{1994}\natexlab{}.
\newblock \showarticletitle{DSC: Scheduling parallel tasks on an unbounded number of processors}.
\newblock \bibinfo{journal}{\emph{IEEE Transactions on parallel and distributed systems}} \bibinfo{volume}{5}, \bibinfo{number}{9} (\bibinfo{year}{1994}), \bibinfo{pages}{951--967}.
\newblock


\bibitem[Yzelman and Bisseling(2012)]%
        {BSPimpl2}
\bibfield{author}{\bibinfo{person}{AN Yzelman} {and} \bibinfo{person}{Rob~H Bisseling}.} \bibinfo{year}{2012}\natexlab{}.
\newblock \showarticletitle{An object-oriented bulk synchronous parallel library for multicore programming}.
\newblock \bibinfo{journal}{\emph{Concurrency and Computation: Practice and Experience}} \bibinfo{volume}{24}, \bibinfo{number}{5} (\bibinfo{year}{2012}), \bibinfo{pages}{533--553}.
\newblock


\bibitem[Yzelman et~al\mbox{.}(2014)]%
        {BSPimpl1}
\bibfield{author}{\bibinfo{person}{AN Yzelman}, \bibinfo{person}{Rob~H Bisseling}, \bibinfo{person}{Dirk Roose}, {and} \bibinfo{person}{Karl Meerbergen}.} \bibinfo{year}{2014}\natexlab{}.
\newblock \showarticletitle{MulticoreBSP for C: a high-performance library for shared-memory parallel programming}.
\newblock \bibinfo{journal}{\emph{International Journal of Parallel Programming}}  \bibinfo{volume}{42} (\bibinfo{year}{2014}), \bibinfo{pages}{619--642}.
\newblock


\bibitem[Yzelman et~al\mbox{.}(2020)]%
        {alp1}
\bibfield{author}{\bibinfo{person}{AN Yzelman}, \bibinfo{person}{D Di~Nardo}, \bibinfo{person}{JM Nash}, {and} \bibinfo{person}{WJ Suijlen}.} \bibinfo{year}{2020}\natexlab{}.
\newblock \showarticletitle{A C++ GraphBLAS: specification, implementation, parallelisation, and evaluation}.
\newblock \bibinfo{journal}{\emph{arXiv preprint arXiv:1906.03196}} (\bibinfo{year}{2020}).
\newblock


\bibitem[Zarebavani et~al\mbox{.}(2022)]%
        {hdagg}
\bibfield{author}{\bibinfo{person}{Behrooz Zarebavani}, \bibinfo{person}{Kazem Cheshmi}, \bibinfo{person}{Bangtian Liu}, \bibinfo{person}{Michelle~Mills Strout}, {and} \bibinfo{person}{Maryam~Mehri Dehnavi}.} \bibinfo{year}{2022}\natexlab{}.
\newblock \showarticletitle{HDagg: hybrid aggregation of loop-carried dependence iterations in sparse matrix computations}. In \bibinfo{booktitle}{\emph{2022 IEEE International Parallel and Distributed Processing Symposium (IPDPS)}}. IEEE, \bibinfo{pages}{1217--1227}.
\newblock


\end{thebibliography}

\newpage

\appendix

\section{Details on our algorithms} \label{app:algos}

This section provides a more detailed discussion of our different algorithms. Recall that the implementations of these algorithms that we used in the experiments (together with our database and the data from the experiments) are available at \cite{folder}, while a more up-to-date version of our scheduling framework can be accessed at \url{https://github.com/Algebraic-Programming/OneStopParallel}. 

Our algorithms can be run via a set of Python scripts, which allow us to configure the files or directories containing the input problems to solve, and several parameters of the solving process. Besides this user interface, the Python scripts also contain the implementations of our ILP-based methods, and are responsible for invoking the CBC ILP solver.

The remaining C++ code (the \texttt{Cilk}, \texttt{ETF} and \texttt{BL-EST} baseline algorithms and their conversion to BSP, the \texttt{BSPg} algorithm, the \texttt{HC} and \texttt{HCcs} local search methods, and the ingredients of the multilevel framework) are available in the file \texttt{simple\_schedulers.cpp}. The only exception is the \texttt{Source} heuristic, which is implemented in a separate file named \texttt{second\_heuristic\_weights\_cluster.cpp}. These two C++ files are automatically compiled and invoked by the Python scripts.

Finally, we provide the Python scripts required for running and evaluating \texttt{HDagg} separately, since these are independent from our own implementations, and they also require the \texttt{HDagg} library as an external dependency \cite{hdagg}.

As a general remark on our algorithms, we also note that some our simpler methods only return an assignment of nodes to processors and supersteps, but not a concrete communication schedule. These algorithms implicitly assume that the corresponding $\Gamma$ is the \textit{lazy communication schedule} defined by the assignment, where every value is sent directly and in the last possible superstep; that is, if a node $v$ has a direct predecessor $u$, and $\pi(v) _{\!} \neq _{\!} \pi(u)$, then we have $(u, \pi(u), \pi(v), \tau(v)-1) \in \Gamma$. In particular, the BSP-converted variants of the baseline algorithms, the \texttt{BSPg} and \texttt{Source} heuristics, and even \texttt{HC} assumes such a lazy schedule, whereas the ILP-based methods and \texttt{HCcs} return a specific, optimized communication schedule. We note that \texttt{ILPinit} also returns a specific $\Gamma$, but this is replaced by the corresponding lazy schedule in our pipeline, since the algorithm is followed by \texttt{HC}, which is only designed to work with lazy communication schedules.

\subsection{Baseline methods}

The main idea of the \texttt{Cilk} work-stealing scheduler was already outlined before: it is a greedy algorithm where every ready task (i.e.\ node with all of its predecessors already computed) is added to the ``ready stack'' of one of the processors. Whenever one of the processors become idle (it finishes computing a node), it takes the topmost ready task from its own stack, and begins the execution of this node. If the processor's stack is empty, it selects another processor with a non-empty stack (uniformly at random among these processors), and starts executing the node at the bottom of this stack, also removing the node from the stack. This ensures that no processor is idle whenever there is a ready node.

We note that originally, \texttt{Cilk} was not defined on DAGs, but instead on processes that spawn further subtasks during execution. In this original setting, each spawned subtask is added to the top of the stack of the processor executing the parent process. In contrast to this original setting, our DAGs do not have parent-child relationships between the given nodes. To adapt the algorithm to DAGs, we use a very similar rule: whenever the execution of the last direct predecessor of a node $v$ is finished, if this happens on processor $p$, then $v$ is added to the top of processor $p$'s stack.

As for \texttt{BL-EST} and \texttt{ETF}, we refer the reader to the work of \cite{SPD} for more details. We note that our baseline implementation for both of these algorithms corresponds to the versions described in \cite{SPD}, which is already extended with the concept of communication volume (but not with latency or NUMA effects, since these are new in our model). In case have NUMA effects, we compute the Earliest Start Time (EST) in these baseline algorithms by considering the average coefficient of the communication cost over all pairs of processors, and multiplying $c(v)$ with this number when considering the required communication steps. Note that an extension of the EST computation with NUMA factors would also be possible, to obtain a version of these baselines that is more specialized towards our more realistic model; however, we leave the investigation of this approach to future work.

We also note that \texttt{Cilk}, \texttt{BL-EST} and \texttt{ETF} assign nodes to concrete points in time. This can be naturally converted into a BSP schedule by sorting it into supersteps iteratively: if some of the nodes are already assigned to supersteps $1$, ..., $(s_{\!}-_{\!}1)$, we can find the earliest point $t$ in time when our classical schedule begins executing a node $v$ such that (i) $v$ is not yet assigned to a superstep, (ii) $v$ has a direct predecessor $v_0$ also not yet assigned to a superstep, and (iii) our schedule has $\pi(v) _{\!} \neq _{\!} \pi(v_0)$. This implies that the next computation phase can last at most until time $t$; hence we assign all nodes that are executed before step $t$ to superstep $s$, and we continue with this procedure for superstep $(s_{\!}+_{\!}1)$.

Finally, \texttt{HDagg} is a more sophisticated algorithmic baseline, which directly develops a BSP-like schedule, sorting the nodes of the DAG into supersteps (called `wavefronts'), and then distributing nodes among processors in each wavefront. The algorithm strives to ensure both a balanced workload among processors in a superstep, and a reduced amount of inter-processor communication between the supersteps. For more details of the algorithm, we refer the reader to \cite{hdagg}. Note that this algorithm was originally developed (and analyzed) for the specific purpose of speeding up SpTRSV computations, i.e.\ solving sparse linear systems defined by a triangular matrix. However, the method is in fact a general DAG scheduling algorithm that can be applied to any computational DAG; it is even analyzed in \cite{hdagg} using DAG terminology. That is, after a topological ordering, the adjacency matrix of any DAG becomes lower triangular, and by artificially adding an edge from each node to itself (i.e.\ setting the main diagonal of the matrix to $1$, thus ensuring it is non-singular), we obtain an SpTRSV problem where the dependency DAG derived from the matrix is identical to our original computational DAG. As such, the schedule returned by \texttt{HDagg} for this matrix directly corresponds to a schedule for our original DAG in the BSP model.

Since the original implementation of \texttt{HDagg} from \cite{hdagg} is openly available, we directly apply this in our experiments: we convert our computational DAG into the required triangular matrix format, we use the example script from the \texttt{HDagg} library to output the returned schedule, and then we evaluate this according to the cost model of our paper.

\subsection{Initialization heuristics} The \texttt{BSPg} heuristics is a greedy method that is designed specifically for developing BSP schedules. Even though the output assigns the nodes to supersteps, the algorithm still considers the concrete time step of starting and finishing each task, like in classical schedulers such as \texttt{Cilk} or \texttt{ETF}; this helps us develop a balanced work cost between the processors in each superstep. When a task is finished, we consider the new tasks $v \in V$ that become ready (i.e.\ have all of their direct predecessors already finished). If $v$ already has a predecessor on multiple processors in the current superstep, then $v$ is added to a general $\text{ready}_{all}$ set of nodes that only become available for all processors in the next superstep. Otherwise, if $v$ only has predecessors on a single processor $p$ in the current superstep, then $v$ is added to a set $\text{ready}_p$ of that processor, signaling that processor $p$ is already allowed to execute $v$ in the current superstep. Whenever a task is finished and some processor $p$ becomes free, they are assigned a new node to execute (based on rules discussed below), if this is still possible without closing the current computation phase, i.e.\ there are still nodes that have all their direct predecessors on $p$ or in earlier supersteps. When half of the processors become idle (they cannot be assigned new nodes), the computation phase is closed, and a following superstep is started, where all the nodes in $\text{ready}_{all}$ are now available to each processor.

When selecting a new node to compute for a processor $p$, we employ the following method (named ChooseNode in our pseudocode). We first attempt to choose from the nodes in $\text{ready}_p$ that are only available for $p$, and if this is empty, we choose from the set $\text{ready}_{all}$ available to all processors. From both sets, we use the same tie-breaking rule that attempts to reduce communication costs, assigning a score metric for each node $v$ in the set, and selecting the node with the highest score. For this score, we consider all direct predecessors $u$ of $v$, and if either $u$ or one of its direct successors is already assigned to $p$, then we increase the score of $v$ by $\frac{c(u)}{\text{outdeg}(u)}$, where $\text{outdeg}(u)$ is the outdegree of $u$. Intuitively, this can be understood as an estimator for the importance and probability of the fact that we can assign $u$ and all its direct successors to $p$, thus not having to communicate $u$ at all. If $c(u)$ is high, then sending $u$ results in a higher cost, and thus more important to avoid; if $\text{outdeg}(u)$ is small, then there is a higher chance that we can indeed achieve this, hence avoiding a communication. As such, this simple score metric for $v$ aims to estimate our opportunities to save communication costs in the future if we assign $v$ to $p$.

We provide a pseudocode for \texttt{BSPg} in Algorithm \ref{alg:BSPg}. Note that $p$ always denotes a processor from $\{ 1, ..., P \}$. We also note that the pseudocode uses in-neighbor and out-neighbor as the short form of direct predecessor and direct successor, respectively.

\begin{algorithm}[h!]
\begin{algorithmic}
    \STATE $superstep \leftarrow 0$
    \STATE $endStep \leftarrow False$
    \STATE $finishTimes \leftarrow \{ 0 \}$
    \STATE \textit{free}$[p] \leftarrow True$ for all $p$
    \WHILE{there are still unassigned nodes}
        \IF{$endStep = True$ \AND $finishTimes = \emptyset$}
            \STATE \textit{ready}$_p \leftarrow \emptyset \:$ for all $p$
            \STATE \textit{ready}$_{all} \leftarrow$ \textit{ready}
            \STATE $superstep \leftarrow superstep+1$
            \STATE $endStep \leftarrow False$
            \STATE $finishTimes \leftarrow \{ 0 \}$
        \ENDIF
        \STATE $t \leftarrow$ earliest time from $finishTimes$
        \FOR{all nodes $v$ that finish at $t$}
            \STATE \textit{free}$[\pi(v)] \leftarrow True$
            \FOR{all out-neighbors $u$ of $v$}
                \IF{$u$'s in-neighbors are all finished}
                    \STATE \textit{ready} $\leftarrow$ \textit{ready} $\cup \, \{ u \}$
                    \IF{$\forall\, (u_0,u) _{\!} \in _{\!} E$ we have either $\pi(u_0) _{\!}= _{\!}\pi(v)$ or $\tau(u_0) _{\!} < _{\!} superstep$}
                        \STATE \textit{ready}$_{\pi(v)}$ $\leftarrow$ \textit{ready}$_{\pi(v)}$ $\cup \, \{ u \}$
                    \ENDIF
                \ENDIF
            \ENDFOR
        \ENDFOR
        \IF{$endStep = False$}
            \WHILE{$\exists \, p$ with \textit{free}$[p] = True$ and \textit{ready}$_{p} \neq \emptyset$ \OR 
                $\qquad \exists \, p$ with \textit{free}$[p] = True$ and \textit{ready}$_{all} \neq \emptyset$}
                \STATE $v \leftarrow \mathbf{ChooseNode}(p)$
                \STATE delete $v$ from \textit{ready}, \textit{ready}$_{p}$ and/or \textit{ready}$_{all}$
                \STATE $\pi(v) \leftarrow p$, $\: \: \tau(v) \leftarrow superstep$
                \STATE $finishTimes \leftarrow finishTimes \cup \{ t+w(v) \}$
                \STATE \textit{free}$[p] \leftarrow False$
            \ENDWHILE
        \ENDIF
        \IF{\textit{ready}$_{all} = \emptyset$ and $\exists \, \frac{P}{2}$ idle processors}
            \STATE \textbf{$endStep \leftarrow True$}
        \ENDIF
    \ENDWHILE
    \caption{\textsf{Summary of the \texttt{BSPg} heuristic}}
    \label{alg:BSPg}
\end{algorithmic}
\end{algorithm}

The \texttt{Source} heuristic is similar in the sense that it begins processing the DAG from the source nodes. In each iteration, it essentially assigns all the current source nodes of the DAG to processors in some way to form the next superstep. It then disregards (``removes'') these source nodes (and their outgoing edges) from the DAG to create the next set of source nodes for the next superstep. In each superstep, the nodes are assigned to processors following a round-robin approach. The first superstep is handled in a special way: the initial source nodes are first organized into clusters, joining pairs of nodes when they have a common direct successor, and then these clusters are assigned to processors in a round-robin way. In all other supersteps, no clustering is done; instead, the source nodes are sorted in decreasing order according to their work weight, and then assigned to processors in this order in a round-robin fashion, to avoid accumulating a too large work cost on any of the processors.

After the round-robin assignment in each superstep, the heuristic also considers the direct successors of all the source nodes, and if for any of them it holds that all its in-neighbors are already assigned to the same processor $p$, then it also assigns this node to processor $p$ in the current superstep. This allows us to avoid creating further supersteps unnecessarily, by slightly relaxing the principle that only the source nodes are assigned in each superstep.

We show a pseudocode for \texttt{Source} in Algorithm \ref{alg:source2}.

\begin{algorithm}[h!]
\begin{algorithmic}
    \STATE $superstep \leftarrow 0$
    \WHILE{there are still unassigned nodes}
        \STATE $sources \, \leftarrow$ all unassigned $v _{\!} \in _{\!} V$ with $\text{indeg}(v) _{\!} = _{\!} 0$
         \STATE $p \leftarrow 1$
        \IF{$superstep=1$}
            \FOR{$v \in sources$}
                \IF{$v$ shares an out-neighbor with another $u \in sources$}
                    \IF{$u$ is already in a cluster}
                        \STATE add $v$ to $u$'s cluster
                    \ELSE
                        \STATE combine $v$ and $u$ into a new cluster
                    \ENDIF
                \ENDIF
            \ENDFOR
            \FOR{all clusters $c$}
                \FOR{all nodes $v \in c$}
                    \STATE $\pi(v) \leftarrow p$, $\: \: \tau(v) \leftarrow superstep$
                    \STATE delete $v$ from $G$
                \ENDFOR
                \STATE $p \leftarrow (p+1) \text{ modulo } P$
            \ENDFOR
        \ELSE
            \STATE sort $sources$ in decreasing order by $w(v)$
            \FOR{$v \in sources$ in order}
                \STATE $\pi(v) \leftarrow p$, $\: \: \tau(v) \leftarrow superstep$
                \STATE delete $v$ from $G$
                \STATE $p \leftarrow (p+1) \text{ modulo } P$
            \ENDFOR
        \ENDIF
        \FOR{all edges $(v,u) _{\!} \in _{\!} E$ with $v \in sources$}
                \IF{all in-neighbors of $u$ are already assigned to $\pi(v)$}
                    \STATE $\pi(u) \leftarrow \pi(v)$, $\: \: \tau(u) \leftarrow superstep$
                \ENDIF
        \ENDFOR
        \STATE $superstep \leftarrow superstep+1$
    \ENDWHILE
    \caption{\textsf{Summary of the \texttt{Source} heuristic}}
    \label{alg:source2}
\end{algorithmic}
\end{algorithm}

Note that both \texttt{BSPg} and \texttt{Source} only define the computation phases of our schedule by assigning nodes to processors and supersteps; the corresponding communication steps for each communication phase are derived in the end according to a lazy communication schedule.

The \texttt{ILPinit} initializer is discussed later, together with the remaining ILP-based methods.

\subsection{Local search}

Our hill climbing methods take an initial solution (BSP schedule) provided by one of the initialization methods, and execute small modifications to this schedule that decrease its total cost, until a local minimum is reached where none of the modifications lead to an improvement (or a predefined time limit is exceeded).

As mentioned before, from each current solution, we consider the modification steps which only change the assignment of single node $v$. In particular, if we currently have $\pi(v)=p$ and $\tau(v)=s$, we consider all the solutions for where we have $\pi(v) \in \{ 1, ..., P \}$ (i.e.\ $v$ on any processor) and $\tau(v) \in \{ (s-1), s, (s+1) \}$ (i.e.\ $v$ in the previous, same, or next superstep), with the assignments for every other node unchanged.

Our \texttt{HC} method also employs several data structures to ensure that the cost of a potential modification can be computed efficiently, and the same data structures can also be updated efficiently after executing a modification. For instance, for each superstep $s$, we keep the work cost and the communication cost of each processor in a sorted set, and for each processor $p$ we keep an external pointer to the entry describing this processor in the set; this allows for a lookup of the maximum in $O(1)$ time, and when updating the cost, a deletion in amortized $O(1)$ and an insertion in logarithmic time.

Furthermore, for each node $v$ and processor $p$, we explicitly store the first superstep $s$ when node $v$ is required on processor $p$ (due to one of $v$'s direct successors); due to the lazy communication schedule, this implies that $v$ will be sent from $\pi(v)$ to $p$ in superstep $(s-1)$ (unless $p=\pi(v)$). This ensures that whenever we consider moving $v$ to a different processor, we can immediately identify the communication steps that are affected by this change, and hence we can efficiently compute the resulting change in the total communication cost.

The list of valid moves in the search space (i.e.\ modification options that still provide a valid BSP schedule) is also computed in a preprocessing phase, and maintained throughout the algorithm; when a node $v$ is moved to another processor and/or superstep, we only need to update the possible move options for $v$ and its direct predecessors/successors.

Together, these data structures ensure that in each step of \texttt{HC}, we simply iterate through the list of valid move options, efficiently check if any of them leads to an improvement, and apply the selected move option, also updating our data structures efficiently.

The communication schedule hill climbing (\texttt{HCcs}) follows a very similar approach. Note that this settings already considers the assignments $\pi$ and $\tau$ to be fixed. For each necessary communication step (when $v$ needs to be sent from $\pi(v)$ to $p$), this assignment naturally defines an earliest and latest communication phase when this transfer can happen: the earliest is $\tau(v)$, and the latest is $(s_0-1)$ for the first superstep $s_0$ that assigns a direct successor of $v$ to processor $p$. Our \texttt{HCcs} method considers alternative communication schedules with $(v, \pi(v), p, s) \in \Gamma$ for different possible $s \in [\tau(v), s_0-1]$. As before, in each step, we consider a modification of selecting a different such $s$ for only one of the required communication steps, with the remaining communication steps in $\Gamma$ unchanged. Note that for simplicity, this setting implicitly assumes that the value of $v$ is directly sent to $p$ from the processor $\pi(v)$ where it was computed. Similarly to \texttt{HC}, this hill climbing also uses sorted sets and external pointers to efficiently maintain the send cost and receive cost of each processor in every superstep.

For both hill climbing methods, there are two possible variants: we either (i) always greedily apply the first modification step that we find which decreases the cost, or (ii) we always consider all the possible modifications from the current solution, and select the one which decreases the cost by the largest amount. Our preliminary experiments have shown that neither of these two approaches is clearly superior to the other in terms of the final schedule found; however, the second method is much more time-consuming, since there are usually numerous possible modification opportunities to analyze from a given solution. Due to this, we have applied the former, greedy variant of the approach in our experiments.

In our experiments, we allow the \texttt{HC}+\texttt{HCcs} methods to run for a time limit of $5$ minutes on our main datasets, and for $30$ minutes on the \texttt{huge} dataset. Given such a time limit, we always allocate $90\%$ of the allowed time to \texttt{HC}, and the remaining $10\%$ to \texttt{HCcs}.

\subsection{ILP-based methods}

Our most sophisticated approach is the ILP-based solving of the scheduling task or some of its subproblems.

The general approach to formulate the scheduling problem as an ILP has already been outlined in \cite{DAGBSP}: our model here corresponds to the submodel called FS by the authors of \cite{DAGBSP}. In this full ILP representation, there is a binary variable \textsc{comp}$_{v, p, s}$ to indicate whether node $v$ is computed on processor $p$ in superstep $s$, and a value \textsc{pres}$_{v, p, s}$ to indicate if the value of $v$ is already available on $p$ by the end of (the computational phase of) superstep $s$. Besides these, the communication steps are represented by binary variables \textsc{comm}$_{v,p_1,p_2,s}$, indicating whether $(v, p_1, p_2, s) \in \Gamma$. These variables allow a natural way to express both the validity conditions of BSP model (precedence constraints and valid communication steps) and the elements of the cost functions with auxiliary variables; see \cite{DAGBSP} for details. For our \texttt{ILPfull} method, we implement the representation above, with only minimal changes: we aggregate some of the (originally separate) linear constraints into a single constraint, and we replace the role of the explicit variables \textsc{pres}$_{v, p, s}$ by the unused variables \textsc{comm}$_{v,p,p,s}$.

In our \texttt{ILPcs} method to optimize the communication schedule, the main idea is similar to that of \texttt{HCcs}: we consider $\pi$ and $\tau$ already fixed from the initial solution, and only try to optimize the necessary communication steps in $\Gamma$. Once again, the assignment defines an earliest superstep $\tau(v)$ and a latest superstep $(s_0-1)$ when we can send a value $v$ from $\pi(v)$ to $p$ if this is required. Hence for each $s \in [\tau(v), s_0-1]$, we define a binary variable \textsc{comm}$_{v, p, s}$ to indicate whether the value is communicated in superstep $s$. Note that similarly to \texttt{HCcs}, this implicitly assumes that the value of $v$ is always sent from $\pi(v)$. Given these variables, the main techniques in the \texttt{ILPcs} formulation are identical to \texttt{ILPfull}: we can use the same approach to ensure with linear constraints that the communication costs are measured properly, and that each communications step is indeed scheduled to some superstep.

The \texttt{ILPpart} and \texttt{ILPinit} formulations are similar to each other in the sense that they both only consider a particular subproblem (a concrete subset of nodes $V_0$ and a concrete interval of supersteps $S_0$), and express this as an ILP. In \texttt{ILPpart}, we begin from a given subset of supersteps $S_0$: we split the BSP supersteps into disjoint intervals (from back to front), and consider the nodes that are currently assigned to the supersteps in $S_0$ in order to obtain our $V_0$. We form these intervals iteratively, always growing the current interval until $|V_0| \cdot |S_0| \cdot P^2$ exceeds $4_{\!}000$; this is because the number of variables in the ILP representation is in the magnitude of $|V_0| \cdot |S_0| \cdot P^2$, so we use this metric to estimate (and limit) the size of the ILP problem obtained. In contrast to this, in case of \texttt{ILPinit}, we begin with a topological ordering of the DAG, and always select the next batch of nodes in this ordering as $V_0$, and set $S_0$ to cover the next $3$ supersteps; once again, the number of nodes in each batch is increased until $|V_0| \cdot 3 \cdot P^2$ exceeds a given threshold ($2_{\!}000$ in this case).

Note that given $V_0$ and $S_0$, the ILP formulation for \texttt{ILPpart} and \texttt{ILPinit} are similar, but differ in the fact that in \texttt{ILPpart}, all the nodes outside of $V_0$ are already assigned to a processor and superstep, whereas in \texttt{ILPinit}, the successors of the nodes in $V_0$ are not assigned yet (hence we disregard them for the optimization).

For the ILP formulation of \texttt{ILPpart} and \texttt{ILPinit}, it is not surprising that the variables \textsc{comp}$_{v, p, s}$ and \textsc{comm}$_{v,p_1,p_2,s}$ are restricted to $v _{\!} \in _{\!} V_0$ and $s _{\!} \in _{\!} S_0$. This already ensures that the ILP problem is drastically smaller than \texttt{ILPfull}, since the number of variables only scales with $|V_0|$ and $|S_0|$, instead of $n$ and the total number of supersteps. However, in these ILPs, there are several further design decisions that allow us to reduce the number of required variables, at the cost of some simple restrictions to the solution space:
\begin{itemize}[topsep=4pt,itemsep=0pt,partopsep=2pt,parsep=7pt]
    \item Some of the nodes $v \in V_0$ might have direct successors $u$ that are scheduled into supersteps after the interval $S_0$. Let \ $\pi(u)=p_1$. Then e.g.\ if we originally had $\pi(v)=p_1$, but the ILP solver sets $\pi(v)=p_2$, then this results in a newly required communication step; on the other hand, if we originally had $\pi(v)=p_2$, but the ILP solver sets $\pi(v)=p_1$, then a previous communication step becomes unnecessary. Note that these communication steps might either be within the superstep interval $S_0$, or after $S_0$. In order to ensure that our superstep formulation only scales with $|S_0|$ (instead of also optimizing communications in supersteps after $S_0$), we choose to ignore the case when $v$ was originally communicated after $S_0$: in this case, reassigning $v$ might make this communication step unnecessary, and hence reduce the cost of an $h$-relation after $S_0$, but we ignore this potential further gain in the ILP objective. On the other hand, whenever $v$ is sent within $S_0$, this is easily captured by our formulation via the variables \textsc{comm}$_{v,p,p',s}$. In particular, in the last communication phase of $S_0$, we only include the variables \textsc{comm}$_{v,p,p',s}$ for processors $p'$ that require the value of $v$ after $S_0$, but for these processors, we also add the constraint that $v$ indeed needs to be present on processor $p'$ by the end of $S_0$. This ensures that our ILP indeed captures the fact that the reassignment comes with newly required communication steps; however, it limits our flexibility by requiring that these new communication steps need to happen until the end of $S_0$.
    \item  Some of the nodes $v \in V_0$ might also have direct predecessors $u$ which are scheduled into supersteps before $S_0$. Similarly to the case of successors, if $v$ is rescheduled to a different processor, then this might result in newly required communication steps or previous communication steps becoming unnecessary. As before, the ignore the potential gains from removable communication steps that were scheduled before $S_0$, in order to ensure that our ILP size only scales with $|S_0|$. Within $S_0$, these communication steps can once again be captured with variables \textsc{comm}$_{u,p_1,p_2,s}$ as before; however, we also need to add such variables for the communication phase of the last superstep before $S_0$. Furthermore, note that the number of such predecessors $u$ can potentially be even larger than $|V_0|$, hence if we add these variables without consideration, they could notably increase the size of the ILP. To avoid this problem in practice, we apply two simplifications. On the one hand, note that $\Gamma$ might ensure that the value of predecessor $u$ is already communicated to several processors $p_0$ before $S_0$; since $u$ is always present on these processors throughout $S_0$, we do not the add the variables \textsc{comm}$_{u,p_1,p_0,s}$ in this case, and we also disregard the precedence constraint $(u,v)$ for \textsc{comp}$_{v, p_0, s}$ (since they are satisfied anyway). On the other hand, as before, we assume that the value of $u$ is directly sent from $\pi(u)$ (which is fixed, since $u$ is computed before $S_0$) to any processor; this removes a factor $P$ from the number of communication variables for $u$. Finally, note that there may also be predecessors $u$ that have both an out-neighbor $v _{\!} \in _{\!} V_0$ and an out-neighbor $u'$ scheduled after $S_0$; in this case, if $u$ is originally sent to $\pi(u')$ during $S_0$, then we also need to add the constraint that the value of $u$ must still be present on $\pi(u')$ by the end of $S_0$ in our ILP solution.
    \item In our communication schedule $\Gamma$, we may also have communication steps $(v, p_1, p_2, s)$ where $v$ was computed before $S_0$, it is only required on $p_2$ after $S_0$, but our $\Gamma$ still just happens to schedule it during the interval $S_0$, even though $v$ has no direct successor in $V_0$. These communication steps have no direct connection to the assignment of the nodes $V_0$, but they still contribute to the communication costs in one of the supersteps of $S_0$. The number of these communication steps can theoretically be as high as $\Theta(n)$; hence if we include them as optimizable options in our ILP formulation, then we may lose the advantage of only scaling with $|V_0|$. As such, we consider these communication steps fixed: the communication costs they incur appear as a pre-computed constant in the send and receive costs of the appropriate processors in the given supersteps, unaffected by our choice of the assignment for $V_0$.
\end{itemize}

As mentioned before, we use the CBC open-source solver \cite{cbc1} for solving the ILP problems described above. Since \texttt{ILPfull} tries to solve the entire problem, we allow a larger time limit of $1$ hour when running this method. \texttt{ILPcs} also looks for a global solution, but in a much more restricted problem, so we set its time limit to $5$ minutes, similarly to \texttt{HC}+\texttt{HCcs}. We select an even shorter time limit for \texttt{ILPpart} and \texttt{ILPinit}: $3$ minutes for the former, $2$ minutes for the latter (since this is supposed to be an even faster heuristic just for initialization). Even this way, \texttt{ILPpart} and \texttt{ILPinit} still dominate the running time of the algorithms in our experiments, since both of these are iterative methods that are repeatedly applied on many different parts of the DAG or the schedule.

\subsection{Multilevel approach}

Our most novel algorithmic contribution is perhaps the application of the multilevel coarsen-solve-refine technique, which is a state-of-the-art approach in hypergraph partitioning tools \cite{multi1, multi2, multi3, DAH}, to the domain of scheduling problems. We develop and analyze a simple variant of this multilevel scheduling approach for the case when a scheduling problem is dominated by very high communication costs, since our other methods often fail to find high-quality solutions in this case.

In the first phase of the multilevel algorithm, our goal is to gradually coarsen the DAG into a significantly smaller representation that captures most of the structure of the original DAG. For this, we repeatedly contract a directed edge of the DAG into a single node, combining all the incoming and outgoing edges of the two nodes. Each such operation reduces the number of nodes by exactly one, so a repeated execution of this procedure allows us to obtain a coarsified DAG representation of any desired size.

In order to have a valid schedule that can be interpreted for each intermediate step of the coarsening procedure, it is critical to ensure that our graph indeed remains a DAG after each contraction step. We ensure this by only selecting edges to contract in each step that satisfy this property. In particular, an edge $(u,v) _{\!} \in _{\!} E$ can be contracted into a single node (without creating a directed cycle) if and only if there is no other directed path in the DAG from $u$ to $v$ apart from the edge $(u,v)$. Note that there are always contractable edges in a DAG: e.g.\ each non-sink node $u$ can be contracted with its out-neighbor $v$ that appears earliest in a topological ordering, since there can be no other directed path from $u$ to $v$.

In our algorithm, we evaluate the contractable edges $(u,v)$ based on two properties: (i) the total work weight $w(u)+w(v)$ that we would obtain after contraction, which should be small to ensure that no large cluster of nodes is forced onto the same processor in the same superstep, and (ii) the communication weight $c(u)$ of the source node, which should preferably be large, since the contraction step implies that the output of $u$ will not require a communication step to be sent to $\pi(v)$, at least not due to this edge. In our implementation, we consider the following simple technique: we sort the list of contractable edges in an increasing order according to $w(u)+w(v)$, and we always select from the first $\frac{1}{3}$ of this list to ensure that we do not merge nodes with large work weight. From this first part of the list, we select the edge with the largest $c(u)$ value.

Note that after the contraction step, we sum up both the work weights and the communication weights of $u$ and $v$ to obtain the weights for the contracted node. This is entirely appropriate for work weights, since these are summed up on the same processor and superstep anyway. For communication weights, this only gives us an estimate (upper bound) on the actual communication requirements: if there is an edge $(u,v)$ in our coarsened DAG, then from (possibly many) original nodes contracted into $u$, maybe only a few had an actual edge to $v$. This means that when a scheduling of the coarse DAG sends the value of $u$ to $v$, and computes the cost of this based on summed weights $c(u)$, then this is only an upper bound on the amount of data that actually has to be communicated.

In our implementation of this coarsening phase, we find the list of all contractable edges in the DAG in the beginning, and we update this list after each contraction step. Note that after contracting $(u,v)$, this does not only involve updating the edges incident to $u$ and $v$: there might be edges arbitrarily far in the DAG that become uncontractable after this step. In particular, if there is another edge $(u',v')$ such that there is a long directed path from $u$ to $v'$, and another one from $u'$ to $v$, then both edges are contractable originally; however, after contracting $(u,v)$, there will be a directed path from $u'$ to $v'$ via the contracted node, so this distant edge $(u',v')$ loses its contractability.

In general, we note that in our implementation, each of the contraction steps are rather time-consuming, due to e.g.\ the need to update the edges described above, and because we always iterate through the list contractable edges to select the one with highest $c(u)$; both of these operations can require $O(|E|)$ time for each contraction step. This is not a problem in our work, since our more complex scheduling methods require significantly more running time anyway. However, it is an interesting question for future work whether we can develop efficient coarsening techniques with significantly smaller time complexity.

Another crucial question regarding the coarsening phase is the optimal amount of coarsening that provides the best schedule in the end. A too coarse DAG might not capture the overall structure well enough, whereas an insufficient amount of coarsening might not provide the advantages of the multilevel approach; in particular, in our setting, it might not remedy the problem of excessively high communication costs (i.e.\ that it is only beneficial to reassign larger clusters of nodes simultaneously). This task of finding the most favorable coarsening ratio is a very challenging and complex problem on its own; investigating this in detail is far beyond the scope of our paper. For our preliminary experiments to validate the multilevel approach, we simply select two specific rates: we coarsen the DAG to $30\%$ and $15\%$ of its original size, run the multilevel approach for both of these cases, and out of the two schedules obtained this way, we select the one with lower cost as the final output of our multilevel scheduler.

We also note that our main question in this coarsening phase was to obtain a significantly coarsened version of the DAG that maintains the acyclic property. This general question has been studied by multiple works in recent years \cite{DAGpart, DAH}, but was mostly evaluated as a partitioning problem in itself, rather than analyzed as a tool for scheduling. The work of \cite{SPD}, as mentioned before, is an exception to this, which has applied one of these acyclic partitioners to further improve state-of-the-art list schedulers. However, in contrast to our iterative coarsening procedure, this approach applies a final acyclic partitioning; as such, it offers no straightforward way to uncoarsen the DAG gradually, and hence there is no opportunity to execute further refinement steps in a multilevel fashion in this case.

The solving phase is the simplest part of the multilevel approach: here we apply the algorithmic pipeline of Figure \ref{fig:pipeline1} on the coarsened DAG. In particular, recall that \texttt{ILPfull} is only applicable on very small DAGs, and \texttt{ILPpart} is also significantly more useful on smaller DAGs; as such, the multilevel approach allows us to also apply these methods more successfully on DAGs that are originally larger.

In the uncoarsening and refinement phase, we gradually undo the contraction steps in a reverse order, thus moving closer and closer to our original DAG. Every time after we execute a given number of uncontraction steps, we \textit{refine} the current schedule. That is, we first project our current schedule (obtained from the solving phase or the previous refinement step) into our current, slightly more uncoarsened DAG: we assign each node to the same processor and same superstep as its contracted counterpart. Note that since the contracted graph was still a DAG, this still produces a valid BSP schedule. However, the newly executed uncontraction steps reveal slightly more of the structure of the original DAG, so we can now fine-tune our schedule towards this. For this, we execute several improvement steps with our local search method (\texttt{HC}) to obtain a slightly more refined variant of our current schedule on this slightly more coarsified version of the DAG.

In our experiments, we choose to refine the schedule after every $5$ uncontraction steps, and we run \texttt{HC} for at most $100$ steps (or until a local minimum is reached). This makes the number of refinement phases proportional to the number of contraction steps; alternatively, one could also opt to only have a fixed number of refinement phases altogether.

We also note that this uncoarsening phase (and hence the whole multilevel approach) can be adapted much more naturally to BSP than to classical models: if nodes are assinged to concrete starting times instead of supersteps, then it is not immediately clear how a schedule on a coarser DAG should be projected to a slightly uncoarsened variant of the same DAG.

Recall that during the refinement steps, we only apply \texttt{HC}, but not \texttt{HCcs}, since the (partially) coarsened DAG often overestimates the required amount of communications (as it merges the weights $c(v)$ in each cluster). Instead, \texttt{HCcs} and \texttt{ILPcs} are applied separately on the original DAG after the uncoarsening has been finished.

Regarding the entire multilevel approach, we note that our implementation is only a preliminary exploration of this idea, and each element of this approach can be further improved to obtain a more advanced version of this algorithm. In particular, analyzing more complex DAG contraction methods, or refinement with more advanced algorithms, or clever methods to estimate the optimal coarsification factor are all promising directions for further improvement. We leave it to future work to investigate these more sophisticated variants of the approach.

Finally, as a side note, we point at that the work of \cite{SPD} also applies the concept of Communication-to-Computation Ratio (CCR) to capture the concept that a scheduling problem is dominated by communication costs, i.e.\ when our multilevel algorithm seems superior to other methods. In the work of \cite{SPD}, CCR was simply defined as the ratio of $\sum_{v \in V} \, c(v)$ and $\sum_{v \in V} \, w(v)$. This metric is not straightforward to generalize to our model with significantly more parameters: multiplying the numerator with $g$ and also the average NUMA coefficient $\frac{\sum \lambda_{p_1, p_2}}{P^2}$ is a rather natural extension, but it is not trivial to include e.g.\ the effect of the parameter $\ell$ in this formula.

\section{Details on the DAG Database} \label{app:database}

This section discusses the details on our computational DAG database. We note that the database itself (coarse grained instances, fine-grained generator, some examples files and tools) are available at \url{https://github.com/Algebraic-Programming/HyperDAG_DB}. The test set of DAGs used in our experiments are available with the remaining supplementary material in \cite{folder}.

As noted before, the DAGs in our database are stored in a hyperDAG format, to be in line with the recent works of \cite{hyperDAG, DAH}. However, this difference is only relevant from a theoretical modelling perspective, to emphasize the fact that even if a node $v$ has multiple out-neighbors on another processor $p$, its output only needs to be sent to $p$ once; due to this, for partitioning problems, it is more adequate to represent the output data of $v$ as a hyperedge, containing $v$ and all its out-neighbors. For our work, this is simply an alternative representation of the precedence constraints in the DAG, and hence it has no effect. In fact, all of our algorithms begin with the simple step of transforming these hyperDAGs back into a regular DAG representation.

\subsection{Coarse-grained DAGs}

In order to obtain the coarse-grained representation of a variety of computations, we considered the GraphBLAS implementation of \cite{alp1}, and extended this with a so-called HyperDAG backend. The goal of this backend is to run simultaneously to a regular GraphBLAS algorithm and gather meta-data during this run, which is then used to generate the representation of the executed computation. For this, the backend considers the various operation primitives that are used as building blocks for the computations in GraphBLAS, and ensures that each of these primitives identifies the inputs and outputs of the given operation, allowing us to reconstruct the structure of the computational DAG. 

This backend automatically allows us to extract the DAG representation of a wide variety of algebraic computations implemented in GraphBLAS. This involves common iterative methods for linear solvers (e.g.\ Conjugate Gradient for positive definite systems, or BiCGStab for general systems), graph algorithms that are naturally expressible in an algebraic form (such as $k$-hop reachability, connected components, or the PageRank algorithm), classical or more advanced methods from machine learning (such as $k$-means, label propagation or sparse neural network inference), and more. For more details on the concrete GraphBLAS computations, we refer the reader to \cite{alp1}. We note that several of the algorithms above are iterative methods; for these, we extract the corresponding computational DAGs both for a predefined small number of iterations (we set this to $3$), and for the case when the algorithm is running until the iterative method converges. 

We note that while larger containers (matrices, vectors) are easy to track within a GraphBLAS algorithm, some simpler data structures, such as scalars, are not trivial to track without introducing extensive changes to the algorithm implementations. Due to this, our extraction from GraphBLAS sometimes provides an incomplete DAG representation, leading to some isolated nodes or smaller isolated components in the resulting DAG. To obtain the test DAGs for our experiments, we simply consider the largest connected component in each of the extracted DAGs; while this does not always cover the entire GraphBLAS algorithm in question, it still represents a subDAG that corresponds to a valid (sub)computation, and anyway captures most of the structure of the whole computation in the majority of cases.

Note that assigning work weights $w(v)$ and communication weights $c(v)$ to the nodes of the extracted DAG is a non-trivial task: while the DAG structure of the computation is fixed for an algorithm, the sizes of the matrices and vectors involved can be arbitrary, based on the inputs in the concrete run. As such, for simplicity, we assign $w(v)=\text{indeg}(v)-1$ to each node $v$ of the DAG, where $\text{indeg}(v)$ is the indegree of $v$: since the operation $v$ represents combining $\text{indeg}(v)$ distinct values, $\text{indeg}(v)-1$ is a reasonable estimation for the workload this requires (consider e.g.\ a summation or a multiplication). The only exception to this is the source nodes of the DAG, where instead of setting $w(v)=0$, we still assign a work cost of $w(v)=1$: while they correspond to inputs of the computation, loading or initializing these values might still require some computational resources. As for communication weights, we uniformly assign a weight of $c(v)=1$ to all nodes in the coarse-grained DAG. We leave it to potential future work to develop a way to assign more accurate weights to the nodes in these extracted DAGs.

\subsection{Fine-grained DAGs}

In order to obtain fine-grained DAG representation of algebraic computations, we have implemented a simple generator tool that synthetically creates and outputs the computational DAG corresponding to a few specific algebraic computations. Each of the four computations depend on a square matrix $A$. In our experiments, $A$ is always defined by a size $N$ (number of rows/columns), and a probability parameter $q$, such that each entry in the matrix is nonzero independently with probability $q$. To construct computations from real-world matrices, the generator also has the option to load input matrices (i.e.\ nonzero patterns) from a file, but this is not used in our experiments.

Given this input matrix, our tool artificially creates the fine-grained computational DAG corresponding to this matrix, with each node describing a simple operation with a few nonzero values (e.g.\ addition or multiplication of scalars). In particular, the generator outputs DAGs corresponding to the following algorithms:
\begin{itemize}[topsep=4pt,itemsep=0pt,partopsep=2pt,parsep=7pt]
    \item \texttt{spmv}: multiplication of a sparse matrix $A$ with a dense vector $u$. An example for this operation is also shown in Figure \ref{fig:coarse_fine}.
    \item \texttt{exp}: an iterative version of \texttt{spmv}, i.e.\ given a sparse matrix $A$ with a dense vector $u$, the naive computation of the vector $A^k _{\!} \cdot _{\!} u$, by executing $k$ distinct \texttt{spmv} operations. This computation is an important building block of many applications.
    \item \texttt{CG}: the well-known conjugate gradient method for finding a numerical solution to a system of linear equations, executed for $k$ iterations.
    \item \texttt{kNN}: in GraphBLAS, this method refers to finding the nodes in a graph that are at most $k$ hops away from a specific node (in contrast to machine learning, where $k$-NN usually covers a different concept). In terms of algebraic computations, this can be represented as to the multiplication of sparse matrix $A$ and a vector $u$ with a single non-zero entry, for $k$ iterations.
\end{itemize}
Besides $N$ and $q$, another defining parameter for the last three algorithms is the number of iterations $k$.

As before, the work weight is set to $1$ for source nodes, and to the indegree of the node minus $1$ for all other nodes; this is indeed realistic for our fine-grained DAGs, since e.g.\ the addition of $4$ scalars indeed requires $3$ addition operations. The communication weights $c(v)$ are again set to $1$ for all nodes.

\subsection{Datasets for our experiments}

To form our datasets, we generate fine-grained DAGs for different $n$ values, and we also strive to produce DAGs of different shape: e.g.\ a higher number of iterations $k$ results in a deeper DAG (where the longest path is longer), in contrast to a smaller $k$, or in contrast to \texttt{spmv} DAGs where no iterations happen at all, and the longest path always has size only $3$.

For the dataset used for the initial training, we create $10$ fine-grained instances, with $n$ ranging from $15$ to $1950$.

For the actual test datasets, we always specified an interval for $n$ first ($[40,80]$ for \texttt{tiny}, $[250,500]$ for \texttt{small}, $[1000,2000]$ for \texttt{medium} and $[5000,10000]$ for \texttt{large}), and we specifically created a fine-grained DAG with each of the $4$ methods in our generator that is approximately in the beginning, middle, and end of this interval. This results in $12$ fine-grained DAGs for the \texttt{tiny} set. For the remaining sets, there is a higher number of possible parameter combinations to create DAGs in the interval; as such, with all the iterative methods that allow more freedom to influence the depth of the DAG (i.e.\ \texttt{exp}, \texttt{cg} and \texttt{kNN}), we create two different DAGs, a deeper and a wider one, for the beginning, middle, and end of each node interval. This results in $6$ fine-grained DAGs with \texttt{exp}, \texttt{cg} and \texttt{kNN}, and $3$ fine-grained DAGs with \texttt{spmv}. Hence altogether, we have $21$ fine-grained DAGs in the \texttt{small}, \texttt{medium} and \texttt{large} datasets.

Besides this, we also add to each dataset the coarse-grained instances in our database where the number of nodes fits into the defined interval. This adds $4$ coarse-grained DAGs to the \texttt{tiny} dataset, and $3$ coarse-grained DAGs to the \texttt{small} dataset.

In order to create the \texttt{huge} dataset, we generate $7$ fine-grained instances with $n _{\!} \in _{\!} [50000, 100000]$: one with \texttt{spmv}, and two with each of the remaining algorithms. We then extend this with $3$ coarse-grained instances where $n$ is in a similar magnitude (one of the instances only has $n_{\!}=_{\!}47023$, the other two are within the interval).

\section{Details on the Experiments} \label{app:exps}

\subsection{Comparison of initializers}

We begin by discussing our preliminary runs on the training set to evaluate the performance of the different initialization methods. With the $10$ training set DAGs and $9$ parameter combinations from $P \in \{ 4, 8, 16\}$ and $g \in \{1, 3, 5\}$, this amounts to $90$ runs altogether.

During these runs, we found that each of our initialization methods can outperform the others. In particular, the best schedule was returned by \texttt{BSPg} in $44$ cases, by \texttt{Source} in $20$ cases and by \texttt{ILPinit} in $26$ cases. Furthermore, we can observe that the relative performance of the heuristics show a strong dependence on the properties of the scheduling problem itself; in particular, on the size of the DAG, the number of processors $P$, and besides these, also on the ``shape'' of the DAG: the shallow DAGs produced by \text{spmv} computations behave rather differently from the rest of the fine-grained DAGs.

\begin{table}
\centering
\caption{Number of times each initialization method is the best, on \texttt{spmv} computations in the training set, separated for different values of $P$.}
\renewcommand{\arraystretch}{1.65}
\begin{tabular}{| c | c | c|}
 $P=4$ & $P=8$ & $P=16$ \\ [0.5ex] 
 \hline\hline
 \makecell{\texttt{Source}: 5 \\ \texttt{BSPg}: 3 \\ \texttt{ILPinit}: 1} & \makecell{\texttt{Source}: 7 \\ \texttt{BSPg}: 2 \\ \texttt{ILPinit}: 0} & \makecell{\texttt{Source}: 7 \\ \texttt{BSPg}: 2 \\ \texttt{ILPinit}: 0} \\ 
 \hline
\end{tabular}
\label{tab:training_spmv}
\end{table}

In particular, Tables \ref{tab:training_spmv} and \ref{tab:training_rest} show the number of times each heuristic turned out to be the most successful, Table \ref{tab:training_spmv} for \text{spmv} computations, and Table \ref{tab:training_rest} for all other fine-grained instances combined. Table \ref{tab:training_spmv} is separated according to $P$, while Table \ref{tab:training_rest} is separated according to $P$ and $n$. There are several straightforward observations from these tables. On the one hand, it is clear that \texttt{Source} is rather effective for the shallow \text{spmv} DAGs, but not very useful otherwise. \texttt{ILPinit} performs well either for very small DAGs, or for very small $P$. Finally, \texttt{BSPg} consistently delivers good results for most of the parameter combinations.

\begin{table}
\centering
\caption{Number of times each initialization method is the best, on \texttt{exp}, \texttt{cg} and \texttt{kNN} computations in the training set, separated for different values of $P$ and different DAG sizes.}
\renewcommand{\arraystretch}{1.85}
\begin{tabular}{c || c | c | c|}
  & $P=4$ & $P=8$ & $P=16$ \\ [0.5ex] 
 \hline\hline
 \makecell{$n \in$ \\ $[15, 120]$} & \makecell{\texttt{ILPinit}: 6 \\ \texttt{Source}: 0 \\ \texttt{BSPg}: 0}  & \makecell{\texttt{ILPinit}: 6 \\ \texttt{Source}: 0 \\ \texttt{BSPg}: 0} & \makecell{\texttt{BSPg}: 4 \\ \texttt{ILPinit}: 1 \\ \texttt{Source}: 1} \\ 
 \hline
 \makecell{$n \in$ \\ $[200, 350]$} & \makecell{\texttt{ILPinit}: 6 \\ \texttt{Source}: 0 \\ \texttt{BSPg}: 0}  & \makecell{\texttt{BSPg}: 6 \\ \texttt{ILPinit}: 0 \\ \texttt{Source}: 0} & \makecell{\texttt{BSPg}: 6 \\ \texttt{ILPinit}: 0 \\ \texttt{Source}: 0} \\
 \hline
 \makecell{$n \in$ \\ $\!\!\![1000, 2000]\!$} & \makecell{\texttt{ILPinit}: 6 \\ \texttt{BSPg}: 3 \\ \texttt{Source}: 0} & \makecell{\texttt{BSPg}: 9 \\ \texttt{ILPinit}: 0 \\ \texttt{Source}: 0} & \makecell{\texttt{BSPg}: 9 \\ \texttt{ILPinit}: 0 \\ \texttt{Source}: 0} \\
 \hline
\end{tabular}
\label{tab:training_rest}
\end{table}

Since \texttt{ILPinit} is a very time-consuming initialization methods, based on these observations, we decide to only run \texttt{ILPinit} for problems with $P_{\!}=_{\!}4$ processors in the experiments; this significantly reduces the running time required for the experiments. Although \texttt{ILPinit} is also often superior when $n$ is small, we do not apply it in this case, since \texttt{ILPfull} and \texttt{ILPiter} can essentially fulfill the same role for these problems. Since both \texttt{BSPg} and \texttt{Source} are very fast heuristics with negligible running time compared to the other elements in our framework, we apply both of them on every input problem, regardless of the parameters.

\subsection{Experiments without NUMA}

The results of our experiments without NUMA effects have already been outlined in Section \ref{sec:exp}. For completeness, here we provide a table with the respective improvements for each combination of $P$, $g$ and dataset, shown in Table \ref{tab:main_app}. Note that each number in the \texttt{tiny}, \texttt{small}, \texttt{medium} and \texttt{large} rows is still the average of $16$, $24$, $21$ and $21$ runs on different DAGs, respectively. This more detailed table reveals some further details compared to Table \ref{tab:base}; for instance, while the cost reduction generally increases with larger $P$, this in fact only holds for the larger datasets, and the effect is in fact the opposite for the \texttt{tiny} DAGs. It also shows that for the most extreme subcase, our scheduler achieves almost a factor $2\times$ improvement with respect to \texttt{HDagg}, and well over a factor $2\times$ improvement with respect to \texttt{Cilk}, even in this non-NUMA setting.

\begin{table*}[t]
    \centering
      \caption{Improvement achieved by our scheduler (without NUMA) for each combination of $g$, $P$ and dataset, with respect to \texttt{Cilk} (first number in cell) and \texttt{HDagg} (second number in cell).}
    \renewcommand{\arraystretch}{1.55}
    \begin{tabular}{c || c | c | c | c | c | c | c | c | c | }
     & \multicolumn{3}{c}{$g=1$} & \multicolumn{3}{c}{$g=3$} & \multicolumn{3}{c}{$g=5$} \\ [0.5ex]
     & $P=4$ & $P=8$ & $P=16$ & $P=4$ & $P=8$ & $P=16$ & $P=4$ & $P=8$ & $P=16$ \\ [0.5ex] 
     \hline\hline
     $\!\!\!$ \texttt{tiny} $\!\!$ & \small $41\% / 34\%$  & \small $ 33\% / 28\%$ & \small $ 20\% / 16\%$ & \small $ 49\% / 43 \%$  & \small $ 40\% / 36 \%$ & \small $ 28\% / 26 \%$ & \small $54 \%  / 49 \%$  & \small $30 \%  / 36 \%$ & \small $ 33 \% / 32 \%$ \\ 
     \hline
     $\!\!\!$ \texttt{small} $\!\!$ & \small $33\% / 23\%$  & \small $41\% / 25\%$ & \small $39\% / 20\%$ & \small $40\% / 28\%$  & \small $46\% / 31\%$ & \small $46\% / 30\%$ & \small $43\% / 30\%$  & \small $46\% / 32\%$ & \small $49\% / 35\%$ \\
     \hline
     $\!\!\!\!\!$ \texttt{medium} $\!\!\!\!\!$ & \small $31\% / 14\%$  & \small $43\% / 17\%$ & \small $53\% / 20\%$ & \small $38\% / 16\%$  & \small $47\% / 20\%$ & \small $56\% / 27\%$ & \small $42\% / 18\%$  & \small $47\% / 20\%$ & \small $58\% / 31\%$ \\
     \hline
     $\!\!\!$ \texttt{large} $\!\!$ & \small $27\% / \,9\%\:$  & \small $41\% / 13\%$ & \small $53\% / 16\%$ & \small $34\% / \,8\%\:$  & \small $46\% / 12\%$ & \small $56\% / 21\%$ & \small $38\% / \,7\%\:$  & \small $46\% / 12\%$ & \small $58\% / 13\%$ \\
     \hline
    \end{tabular}
  \label{tab:main_app}
\end{table*}

\begin{table*}[t]
    \centering
    \caption{Ratio of costs achieved by our algorithms (similarly to Figure \ref{fig:base_diag}), for $g=5$, on the different datasets.}
    \renewcommand{\arraystretch}{1.55}
    \begin{tabular}{c || c | c | c | c | c | c | c | c | }
     & \texttt{BL-EST} & \texttt{ETF} & \texttt{Cilk} & \texttt{HDagg} & \texttt{Init} & \texttt{HCcs} & \texttt{ILPpart} & \texttt{ILPcs} \\ [0.5ex] 
     \hline\hline
     $\!\!$ \texttt{tiny} & $1.126$ & $0.883$ & $\:1\:$ & $0.943$ & $0.728$ & $0.619$ & $0.57$ & $0.569$ \\ 
     \hline
     $\!\!$ \texttt{small} & $1.54$ & $1.073$ & $\:1\:$ & $0.791$ & $0.66$ & $0.579$ & $0.556$ & $0.539$ \\
     \hline
     $\!\!$ \texttt{medium} & $1.896$ & $1.254$ & $\:1\:$ & $0.658$ & $0.592$ & $0.542$ & $0.529$ & $0.506$ \\
     \hline
     $\!\!$ \texttt{large} & $2.142$ & $1.517$ & $\:1\:$ & $0.609$ & $0.591$ & $0.547$ & $0.542$ & $0.521$ \\
     \hline
    \end{tabular}
  \label{tab:main_app2}
\end{table*}

In order to better understand the different ILP methods, we show the performance of each algorithm on the datasets for $g _{\!}=_{\!}5$ (and all of $P _{\!} \in _{\!} \{ 4, 8, 16\}$) in Table \ref{tab:main_app2}, with the ratios normalized to \texttt{Cilk} as in our figures before. The \texttt{ILPpart} column shows the relative cost of the schedule after running \texttt{ILPfull} and \texttt{ILPpart}, whereas \texttt{ILPcs} shows the final schedule after also running \texttt{ILPcs}. The table shows that in the \texttt{tiny} dataset, \texttt{ILPfull}/\texttt{ILPpart} has a significant effect, decreasing the mean ratio from $0.619$ to $0.569$, which amounts to a $8\%$ cost decrease from $0.619$. As $n$ grows larger, the improvement achieved by these methods becomes less significant. In contrast to this, \texttt{ILPcs} only achieves a minimal improvement on the smaller dataset, but it is more impressive when $n$ is larger: it decreases the ratio from $0.529$ to $0.506$ on \texttt{medium} and from $0.542$ to $0.521$ on \texttt{large} ($4\%$ improvement in both cases). This suggests that \texttt{ILPpart} and \texttt{ILPcs} excel in different situations, and hence they are both valuable ingredients in our scheduler.

The table also contains our other academic baselines, \texttt{ETF} and \texttt{BL-EST}. The data shows that as $n$ grows, even \texttt{Cilk} becomes more superior to both \texttt{ETF} and \texttt{BL-EST}. Also, both \texttt{ETF} and \texttt{BL-EST} are significantly outperformed by \texttt{HDagg}, except for a single case: \texttt{ETF} performs better than \texttt{HDagg} on the \texttt{tiny} dataset. Since \texttt{ETF} is the strongest baseline altogether for this specific dataset, for completeness, we also show the cost improvement of our scheduler on the \texttt{tiny} dataset compared to \texttt{ETF} in Table \ref{tab:etf}. The table shows that our scheduler is also consistently superior to \texttt{ETF} by a significant margin.

\begin{table}
\centering
\caption{Cost reduction achieved by our scheduler compared to \texttt{ETF} on the \texttt{tiny} dataset, for each combination of $g$ and $P$, without NUMA.}
\renewcommand{\arraystretch}{1.65}
\begin{tabular}{c || c | c | c|}
  & $g=1$ & $g=3$ & $g=5$ \\ [0.5ex] 
 \hline\hline
 $P=4$ & $38\%$  & $43\%$ & $46\%$ \\ 
 \hline
 $P=8$ & $33\%$ & $31\%$ & $32\%$ \\
 \hline
 $P=16$ & $22\%$ & $27\%$ & $28\%$ \\
 \hline
\end{tabular}
\label{tab:etf}
\end{table}

\subsection{The role of latency}

So far, we always had a fixed choice of $\ell=5$ in our scheduling problems. As such, we also run a small experiment to investigate the effect of the parameter $\ell$ on our schedules. For this, we consider the \texttt{medium} dataset, with a choice of $g_{\!}=_{\!}1$ (to ensure that communication costs are dominated by $\ell$) and $P_{\!}=_{\!}8$. We investigate different values for the latency in this setting: $\ell _{\!} \in _{\!} \{ 2, 5, 10, 20\}$.

We show the improvement achieved by our scheduler for each of these cases in Table \ref{tab:latency}. The table shows that similarly to $g$, the improvement increases for larger values of $\ell$. However, in contrast to $g$, the latency needs to be set to much larger numerical values in order for this tendency to become clearly noticeable.

\begin{table}
\centering
\caption{Cost reduction achieved by our scheduler for different values of $\ell$, on the \texttt{medium} dataset for $g_{\!}=_{\!}1$ and $P_{\!}=_{\!}8$ (compared to \texttt{Cilk}/\texttt{HDagg}).}
\renewcommand{\arraystretch}{1.65}
\begin{tabular}{| c | c | c | c|}
 $\ell=2$ & $\ell=5$ & $\ell=10$ & $\ell=20$ \\ [0.5ex] 
 \hline\hline
 $38\% \: / \: 16\%$ & $43\% \: / \: 17\%$  & $50\% \: / \: 19\%$ & $58\% \: / \: 21\%$ \\ 
 \hline
\end{tabular}
\label{tab:latency}
\end{table}

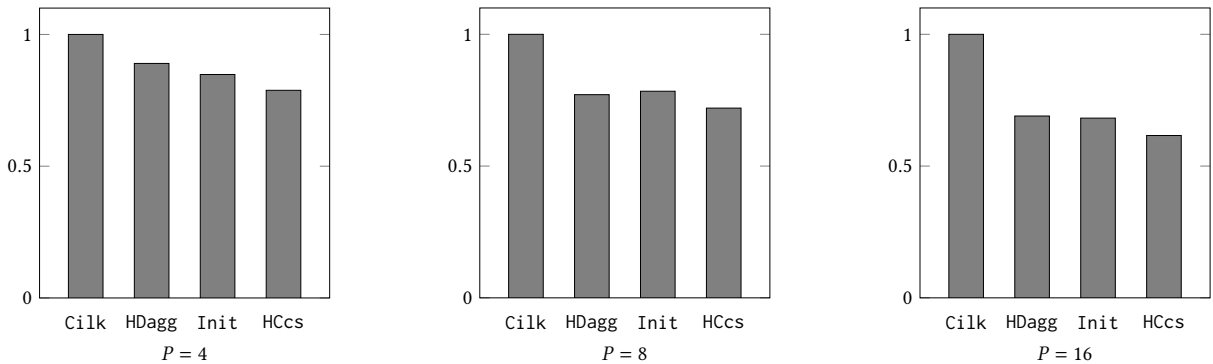
\begin{figure*}[ht]
    \centering
  \begin{minipage}[b]{0.25\textwidth}
    \resizebox{1.0\textwidth}{!}{
    \begin{tikzpicture} 
    \begin{axis}
        [ybar,
        ymin=0,ymax=1.1,
        bar width=15pt,
        height=6cm, width=6cm,
        enlarge x limits={abs=20pt},
        symbolic x coords={\texttt{Cilk}, \texttt{HDagg}, \texttt{Init}, \texttt{HCcs}},
        xlabel={$P=4$},
        major x tick style = transparent]
    \addplot[draw=black, fill=gray] coordinates {
        (\texttt{Cilk},1) 
        (\texttt{HDagg},0.89) 
        (\texttt{Init},0.848) 
        (\texttt{HCcs},0.788)
    };
    \end{axis}
    \end{tikzpicture}}
  \end{minipage}
  \hspace{0.07\textwidth}
    \begin{minipage}[b]{0.25\textwidth}
    \resizebox{1.0\textwidth}{!}{
    \begin{tikzpicture} 
    \begin{axis}
        [ybar,
        ymin=0,ymax=1.1,
        bar width=15pt,
        height=6cm, width=6cm,
        enlarge x limits={abs=20pt},
        symbolic x coords={\texttt{Cilk}, \texttt{HDagg}, \texttt{Init}, \texttt{HCcs}},
        xlabel={$P=8$},
        major x tick style = transparent]
    \addplot[draw=black, fill=gray] coordinates {
        (\texttt{Cilk},1) 
        (\texttt{HDagg},0.771) 
        (\texttt{Init},0.784) 
        (\texttt{HCcs},0.72)
    };
    \end{axis}
    \end{tikzpicture}}
  \end{minipage}
  \hspace{0.07\textwidth}
    \begin{minipage}[b]{0.25\textwidth}
    \resizebox{1.0\textwidth}{!}{
    \begin{tikzpicture} 
    \begin{axis}
        [ybar,
        ymin=0,ymax=1.1,
        bar width=15pt,
        height=6cm, width=6cm,
        enlarge x limits={abs=20pt},
        symbolic x coords={\texttt{Cilk}, \texttt{HDagg}, \texttt{Init}, \texttt{HCcs}},
        xlabel={$P=16$},
        major x tick style = transparent]
    \addplot[draw=black, fill=gray] coordinates {
        (\texttt{Cilk},1) 
        (\texttt{HDagg},0.69) 
        (\texttt{Init},0.682) 
        (\texttt{HCcs},0.616)
    };
    \end{axis}
    \end{tikzpicture}}
  \end{minipage}
  \hfill
  \caption{Improvement achieved by the initialization methods and the local search algorithm on the \texttt{huge} dataset, separated according to the value of $P$.}
  \label{fig:huge}
\end{figure*}

\subsection{Experiments with NUMA}

In our experiments with NUMA effects, we only considered $P_{\!}=_{\!}8$ and $P_{\!}=_{\!}16$ for the number of processors, since even as a binary tree, $P_{\!}=_{\!}4$ gives a very shallow hierarchy with only two levels. Similarly, we only considered $g_{\!}=_{\!}1$, since with $P_{\!}=_{\!}16$ and $\Delta_{\!}=_{\!}4$, this already gives a coefficient as high as $\Delta^3_{\!}=_{\!}64$ between pairs of processors that are only connected over the highest level of the hierarchy. As before, we set $\ell_{\!}=_{\!}5$. For the NUMA multiplier, we consider the values $\Delta _{\!} \in _{\!} \{ 2, 3, 4\}$. Generally, as $n$ grows larger, the improvement with respect to \texttt{HDagg} typically get smaller, whereas the improvement with respect to \texttt{Cilk} does not show such a clear pattern in terms of $n$.

\begin{table*}[t]
    \centering
      \caption{Improvement achieved by our scheduler (with NUMA) for each combination of $P$, $\Delta$ and dataset, with respect to \texttt{Cilk} (first number) and \texttt{HDagg} (second number), for a fixed choice of $g=1$ (and $\ell=5$).}
    \renewcommand{\arraystretch}{1.55}
    \begin{tabular}{c || c | c | c | c | c | c | }
     & \multicolumn{3}{c}{$P=8$} & \multicolumn{3}{c}{$P=16$} \\ [0.5ex]
     & $\Delta=2$ & $\Delta=3$ & $\Delta=4$ & $\Delta=2$ & $\Delta=3$ & $\Delta=4$ \\ [0.5ex] 
     \hline\hline
     $\!$ \texttt{tiny}  & $43\% \: / \: 39\%$ & $57\% \: / \: 54\%$ & $66\% \: / \: 64\%$ & $45\% \: / \: 45\%$  & $68\% \: / \: 68\%$ & $77\% \: / \: 78\%$ \\ 
     \hline
     $\!$ \texttt{small} & $48\% \: / \: 31\%$ & $55\% \: / \: 40\%$ & $60\% \: / \: 47\%$ & $55\% \: / \: 38\%$  & $66\% \: / \: 52\%$ & $71\% \: / \: 59\%$ \\
     \hline
     $\!\!$ \texttt{medium} $\!$ & $50\% \: / \: 23\%$ & $55\% \: / \: 30\%$ & $58\% \: / \: 35\%$ & $61\% \: / \: 34\%$  & $67\% \: / \: 44\%$ & $69\% \: / \: 49\%$ \\
     \hline
     $\!$ \texttt{large} & $49\% \: / \: 14\%$ & $54\% \: / \: 18\%$ & $57\% \: / \: 20\%$ & $61\% \: / \: 28\%$  & $67\% \: / \: 38\%$ & $69\% \: / \: 42\%$ \\
     \hline
    \end{tabular}
  \label{tab:numa1}
\end{table*}

The improvements achieved by our scheduler, separated according to $P$, $\Delta$ and the dataset, are shown in Table \ref{tab:numa1}. The table confirms that the improvement achieved by our scheduler grows with a larger $P$ and $\Delta$, and this general rule holds consistently for all cases in all of the datasets, i.e.\ regardless of $n$. Compared to \texttt{Cilk}, the difference between the parameter combinations is most significant in the \texttt{tiny} dataset: the smallest and largest improvement is both achieved in this dataset.

Recall that the same results are aggregated over the different dataset (for each $P$ and $\Delta$) in Table \ref{tab:numa2}. Finally, combined over all datasets and all values of $P$ and $\Delta$, the mean improvement is $60\%$ and $43\%$ compared to \texttt{Cilk} and \texttt{HDagg}, respectively. We note that \texttt{ETF} and \texttt{BL-EST} are also clearly inferior in this setting with NUMA effects: on all datasets combined, they returns a schedule that is on average $4\%$ and $41\%$ worse, respectively, than even \texttt{Cilk}.

\subsection{The \texttt{huge} dataset}

Our results on the \texttt{huge} dataset without NUMA effects are shown in Table \ref{tab:huge}, for separate values of $P$ and $g$. As mentioned before in Section \ref{sec:exp}, the improvement compared to \texttt{Cilk} ranges from $15\%$ to $41\%$, and the improvement to \texttt{HDagg} ranges from $6\%$ to $13\%$.

To also see the amount of improvement we can attribute to the initializers and to \texttt{HC}+\texttt{HCcs}, we illustrate the improvement in Figure \ref{fig:huge}. This shows that the initializers achieve a $15\%$, $22\%$ and $32\%$ cost reduction compared to \texttt{Cilk} (for $P=4,\, 8, \, 16$, respectively). The local search methods then further improve this by $7\%$, $8\%$ and $10\%$ (for $P=4,\, 8, \, 16$, respectively) compared to \texttt{Init}. Hence even for these much larger DAGs, without applying our ILP-based methods, our schedulers achieve a rather significant improvement to \texttt{Cilk}, especially for higher $P$ values. When compared to \texttt{HDagg}, we see that \texttt{Init} is approximately on par with this baseline, and altogether, \texttt{Init}+\texttt{HC}+\texttt{HCcs} provides an improvement of $11\%$, $7\%$ and $11\%$ (for $P=4,\, 8, \, 16$, respectively). We leave it to future work to improve the scaling of our algorithms, to ensure that they are also superior to \texttt{HDagg} by a more significant margin for DAGs of this size.

\begin{table}
\centering
\caption{Cost reduction achieved by \texttt{Init}+\texttt{HC}+\texttt{HCcs} on the \texttt{huge} dataset without NUMA, compared to \texttt{Cilk}/\texttt{HDagg}.}
\renewcommand{\arraystretch}{1.65}
\begin{tabular}{c || c | c | c|}
  & $g=1$ & $g=3$ & $g=5$ \\ [0.5ex] 
 \hline\hline
 $P=4$ & $15\% \: / \: 9\%$  & $22\% \: / \: 12\%$ & $26\% \: / \: 13\%$ \\ 
 \hline
 $P=8$ & $24\% \: / \: 7\%$ & $30\% \: / \: \, 6\:\%$ & $30\% \: / \: \, 7\%\:$ \\
 \hline
 $P=16$ & $35\% \: / \: 9\%$ & $39\% \: / \: 11\%$ & $41\% \: / \: 13\%$ \\
 \hline
\end{tabular}
\label{tab:huge}
\end{table}

The results on the \texttt{huge} dataset with NUMA effects are mostly similar; these are shown in Table \ref{tab:huge_numa}, separated according to $P$ and $\Delta$. The improvement here ranges from $30\%$ to $48\%$ compared to \texttt{Cilk}, and $7\%$ to $21\%$ compared to \texttt{HDagg}. As such, similarly to the smaller dataset, the improvements become larger when we have NUMA effects; however, the difference between our scheduler and \texttt{HDagg} still remains relatively small in several of the cases.

\begin{table}[t]
\centering
\caption{Cost reduction achieved by \texttt{Init}+\texttt{HC}+\texttt{HCcs} on the \texttt{huge} dataset with NUMA, compared to \texttt{Cilk}/\texttt{HDagg}.}
\renewcommand{\arraystretch}{1.65}
\begin{tabular}{c || c | c | c|}
  & $\Delta=2$ & $\Delta=3$ & $\Delta=4$ \\ [0.5ex] 
 \hline\hline
 $P=8$ & $30\% \: / \: \,7\%\:$ & $34\% \: / \: \,7\%\:$ & $37\% \: / \: \,7\%\:$ \\
 \hline
 $P=16$ & $41\% \: / \: 12\%$ & $45\% \: / \: 16\%$ & $48\% \: / \: 21\%$ \\
 \hline
\end{tabular}
\label{tab:huge_numa}
\end{table}

\subsection{Multilevel scheduling}

We primarily experiment with our multilevel scheduler in the settings where the communication costs are high; in our case, this mostly means the setting with NUMA effects. In our experiments, we only consider the multilevel scheduler on the \texttt{small}, \texttt{medium} and \texttt{large} datasets, since coarsening the DAGs in \texttt{tiny} would lead to an absurdly small DAG representation with as few as $6$ nodes in the most extreme case.

\begin{table*}[t!]
    \centering
      \caption{Cost reduction achieved by the multilevel scheduler in case of NUMA with respect to \texttt{Cilk}/\texttt{HDagg}, for each combination of $P$ and $\Delta$, on the combination of all datasets except \texttt{tiny}, for a fixed choice of $g=1$ (and $\ell=5$). The rows \texttt{C}$_{15}$ and \texttt{C}$_{30}$ show the result obtained when running the multilevel algorithm with a coarsification factor of $15\%$ and $30\%$, respectively, while the \texttt{C}$_{opt}$ row denotes the variant when we run both of these algorithms, and select the schedule of lower cost from the two outputs.}
    \renewcommand{\arraystretch}{1.55}
    \begin{tabular}{c || c | c | c | c | c | c | }
     & \multicolumn{3}{c}{$P=8$} & \multicolumn{3}{c}{$P=16$} \\ [0.5ex]
     & $\Delta=2$ & $\Delta=3$ & $\Delta=4$ & $\Delta=2$ & $\Delta=3$ & $\Delta=4$ \\ [0.5ex] 
     \hline\hline
     $\!$ \texttt{C}$_{15}$  & $31\% \: / \, -3\%$ & $48\% \: / \: 21\%$ & $63\% \: / \: 41\%$ & $47\% \: / \: 14\%$  & $73\% \: / \: 56\%$ & $85\% \: / \: 75\%$ \\ 
     \hline
     $\!$ \texttt{C}$_{30}$ & $39\% \: / \: \, 9\% \:$ & $54\% \: / \: 29\%$ & $64\% \: / \: 44\%$ & $53\% \: / \: 24\%$  & $74\% \: / \: 58\%$ & $85\% \: / \: 75\%$ \\
     \hline
     $\!\!$ \texttt{C}$_{opt}$ & $40\% \: / \: 10\%$ & $56\% \: / \: 32\%$ & $67\% \: / \: 48\%$ & $54\% \: / \: 26\%$  & $76\% \: / \: 61\%$ & $87\% \: / \: 79\%$ \\
     \hline
    \end{tabular}
  \label{tab:ml1}
\end{table*}

\begin{table*}[t!]
    \centering
      \caption{Improvement factor achieved by the multilevel scheduler with respect to our base scheduler (the framework of Figure \ref{fig:pipeline1}), in the same setting as described in Table \ref{tab:ml1}.}
    \renewcommand{\arraystretch}{1.55}
    \begin{tabular}{c || c | c | c | c | c | c | }
     & \multicolumn{3}{c}{$P=8$} & \multicolumn{3}{c}{$P=16$} \\ [0.5ex]
     & $\Delta=2$ & $\Delta=3$ & $\Delta=4$ & $\Delta=2$ & $\Delta=3$ & $\Delta=4$ \\ [0.5ex] 
     \hline\hline
     $\!$ \texttt{C}$_{15}$  & $1.353 $ & $ 1.136 $ & $ 0.912 $ & $ 1.291 $  & $ 0.813 $ & $ 0.506 $ \\ 
     \hline
     $\!$ \texttt{C}$_{30}$ & $1.195 $ & $ 1.014 $ & $ 0.871 $ & $ 1.141 $  & $ 0.774 $ & $ 0.502 $ \\ 
     \hline
     $\!\!$ \texttt{C}$_{opt}$ & $1.179 $ & $ 0.979$ & $ 0.812 $ & $ 1.122 $  & $ 0.711 $ & $ 0.429 $ \\ 
     \hline
    \end{tabular}
  \label{tab:ml2}
\end{table*}

Note that if the communication costs are very high in general (such as with NUMA), then finding a good schedule becomes a very challenging task: intuitively, it is only beneficial to assign two parts of the DAG to separate processors if the relative size of these parts is much larger than the number of edges going between them, otherwise the corresponding communication steps result in more extra cost than what we gain from the parallel execution of the subtasks. In particular, when we have high NUMA costs in our model, it can happen in our experiments that the best solution found by both the baseline methods and our scheduler is in fact more costly than the trivial solution of assigning the entire DAG to a single processor and single superstep. This clearly shows that the schedulers (both the baselines and ours) fail to find any reasonable schedule in this case. In particular, with very high communication cost parameters, it is not even clear at first whether a solution of lower cost even exists (just not found by our schedulers), or whether the trivial solution mentioned above is in fact the optimal schedule in these problems.

Our multilevel algorithm answers this question, demonstrating that an approach designed especially for this communication-heavy case can find a solution with lower than the trivial cost even in this very challenging setting. In particular, over all experiments with NUMA costs ($P _{\!} \in _{\!} \{ 8, 16\}$, $\Delta _{\!} \in _{\!} \{ 2, 3, 4\}$), there were as many as $114$ out of $396$ cases where our base scheduler (and the baselines) could not find a solution with lower than trivial cost; however, with the application of the multilevel algorithm, the number of such cases was only $8$ out of $396$. These numbers are understood over all datasets except \texttt{tiny} (where multilevel scheduling was not applied, and hence it cannot be included in this comparison). This shows that even our relatively simple implementation of this multilevel idea can indeed very nicely fill the specialist role of scheduling these kind of problems.

The concrete improvement factors achieved by our multilevel algorithm (separately for $P _{\!} \in _{\!} \{ 8, 16\}$, $\Delta _{\!} \in _{\!} \{ 2, 3, 4\}$) are shown in Table \ref{tab:ml1} with respect to the baselines, and in Table \ref{tab:ml2} with respect to our (non-multilevel) scheduling framework. The tables show that compared to the baselines, the multilevel approach can achieve a cost reduction of up to $87\%$ and $79\%$ in the most extreme case (with respect to \texttt{Cilk}/\texttt{HDagg}), which is more than a factor $7.7\times$ improvement from \texttt{Cilk} and more than a factor $4.7\times$ from \texttt{HDagg}. When compared to our base scheduling framework, the multilevel approach clearly remains inferior when $\Delta_{\!}=_{\!}2$, it is approximately equally strong when $\Delta_{\!}=_{\!}2$ and $P_{\!}=_{\!}8$, and is clearly superior when we have $\Delta_{\!}=_{\!}3$ and $P_{\!}=_{\!}16$, or $\Delta_{\!}=_{\!}4$. For the most extreme case of $\Delta_{\!}=_{\!}4$ and $P_{\!}=_{\!}16$, it provides another more than $2\times$ improvement compared to the base scheduling framework.

Note that the tables are split into multiple rows based on whether we use the approach with a coarsification factor of $0.15$ or $0.3$, or whether we run both and select the better out of the two schedules. The table shows that applying and comparing both coarsifications indeed leads to some further improvement over the single-ratio approach. In case if we prefer running the multilevel approach with only a single ratio, $0.3$ seems to be the clearly superior choice from the two.

In contrast to these cases, when we apply the multilevel scheduler to settings with lower communication costs, it typically provides weaker schedules than our base scheduling approach. This was already indicated before by the case of $\Delta_{\!}=_{\!}2$ in Table \ref{tab:ml2}. In our setting without NUMA, the difference is even larger: for instance, the mean ratio of the multilevel algorithm to our base scheduler is $1.246$ for a choice of $P_{\!}=_{\!}8$ and $g_{\!}=_{\!}3$, and $1.515$ for a choice of $P_{\!}=_{\!}16$ and $g_{\!}=_{\!}1$ (still understood over all datasets except \texttt{tiny}). While these ratios imply that the multilevel approach is still somewhat better than \texttt{Cilk} (and even marginally better than \texttt{HDagg} in the $P_{\!}=_{\!}8$, $g_{\!}=_{\!}3$ case), altogether, it is clearly inferior to our base scheduler. This suggests that our current implementation of the multilevel method can indeed be understood as a specialized tool for the case when communication costs are really dominant in the scheduling problem.

However, we believe that with further polishing, the multilevel approach can be improved to find a favorable coarsification ratio by itself (intuitively, to decide if coarsification is even necessary, or in a more advanced case, on which parts of the DAG it is necessary), and hence it can become a technique that provides high-quality schedules on all problems, regardless of input parameters. We believe that this is one of the most promising directions for future work in this topic.

\end{document}